\documentclass[11pt]{article}
\usepackage[utf8]{inputenc}
\usepackage[numbers]{natbib}
\usepackage{amssymb,amsthm}
\usepackage{xcolor}
\usepackage{hyperref}
\usepackage{amsxtra}
\usepackage{mathtools}
\usepackage{txfonts}
\mathtoolsset{showonlyrefs=true}
\usepackage{booktabs}
\usepackage[noblocks,affil-it]{authblk}
\usepackage[margin=1em]{subfig}
\usepackage{todonotes} 

\newtheorem{theorem}{Theorem}[section]
\newtheorem{corollary}{Corollary}

\newtheorem{lemma}[theorem]{Lemma}
\newtheorem{proposition}{Proposition}

\theoremstyle{definition}
\newtheorem{definition}[theorem]{Definition}
\newtheorem{remark}{Remark}

\newcommand{\Cc}{\mathbb{C}}
\newcommand{\Nn}{\mathbb{N}}
\newcommand{\Rr}{\mathbb{R}}
\newcommand{\Ss}{\mathbb{S}}
\newcommand{\Zz}{\mathbb{Z}}
\newcommand{\Dd}{\mathfrak{D}}
\newcommand{\GG}{\mathfrak{g}}
\newcommand{\SL}{\mathfrak{sl}}
\newcommand{\SU}{\mathfrak{su}}
\newcommand{\UU}{\mathfrak{u}}
\DeclareMathOperator{\Tr}{Tr}

\title{An algebraic approach to the spontaneous formation of spherical jets}
\date{}
\author{}
\begin{document}
\maketitle

% Enter the first author's name and address:
\centerline{\scshape Milo Viviani}
\medskip
{\footnotesize
% please put the address of the first author
 \centerline{CRM Ennio De Giorgi}
   \centerline{Collegio Puteano, Scuola Normale Superiore
Piazza dei Cavalieri, 3}
   \centerline{ Pisa, I-56100, Italy}
} % Do not forget to end the {\footnotesize by the sign }

%\medskip
%
%\centerline{\scshape First-name2 last-name2 and First-name3
%last-name3}
%\medskip
%{\footnotesize
% % please put the address of the second  and third author
% \centerline{ First line of the address of the second author}
%   \centerline{Other lines}
%   \centerline{Springfield, MO 65810, USA}
%}

\bigskip

% The name of the associate editor will be entered by an editorial staff
% "Communicated by the associate editor name" is not needed for special issue.

%The abstract of your paper
\begin{abstract}
The global structure of the atmosphere and the oceans is a continuous source of intriguing challenges in geophysical fluid dynamics (GFD).
Among these, jets are determinant in the air and water circulation around the Earth.
In the last fifty years, thanks to the development of more and more precise and extensive observations, it has been possible to study in detail the atmospheric formations of the giant-gas planets in the solar system. 
For those planets, jets are the dominant large scale structure.
Starting from the 70s, various theories combining observations and mathematical models have been proposed in order to describe their formation and stability.
In this paper, we propose a purely algebraic approach to describe the spontaneous formation of jets on a spherical domain.
Analysing the algebraic properties of the 2D Euler equations, we give a characterization of the different jets' structures.
The calculations are performed starting from the discrete Zeitlin model of the Euler equations.
For this model, the classification of the jets' structures can be precisely described in terms of reductive Lie algebras decomposition.
The discrete framework provides a simple tool for analysing both from a theoretical and and a numerical perspective the jets' formation.
Furthermore, it allows to extend the results to the original Euler equations.
\end{abstract}

%The title of your section 1
\section{Introduction}

Fluid dynamics on planetary scales has a variety of characteristic features.
Direct observations of oceans and atmosphere and more recently non-terrestrial atmospheres have revealed different complex structures.
Among those, jets are some of the most distinguished ones, especially on giant planets like Jupiter or Saturn.
First observed by Galileo Galilei in 1610, Jupiter's atmosphere posses a banded structure of visible clouds.
Thanks to the Pioneer probes in the early 70s, it has been possible to notice that the bands on Jupiter have a non-homogeneous dynamics, dominated by zonal circulations with different velocities.
Hence, the bands on the Jupiter's atmosphere represent several spherical jets.

In 1973, Rhines \cite{Rhi1973} determined a spatial quantity, now called \textit{Rhines scale},
\[
L_{\beta} = \pi\left(\dfrac{2U}{\beta}\right)^{1/2},
\]
where $U$ is the rms velocity (or equivalently the square root of the energy density \cite{MaWa2006}) of the fluid and $\beta$ the northward gradient of the Coriolis vorticity, at which the jets of width $L_\beta$ occur. 
He also conjectured that the concentration of energy at the Rhines scale would have determined a long lasting formation of alternating zonal jets.
Soon later in 1975, G. Williams \cite{Wil1975} numerically proved the conjecure of Rhines, showing that the jets appearance can be understood as the interaction of the 2D turbulence cascade and the Rossby-Haurwitz wave propagation, due to the rotation of the planet.
Furthermore, Williams was also able to show that the zonal formations might contain some trapped vortices, similar to Jupiter's Red Spot.
Since then, various numerical simulations have tried to verify the results obtained by Williams using different fluid models.
For a complete and recent collection of results, see the book \cite{GalRea2019}.

In the atmosphere and ocean's studies, it is common to assume the fluid to be two-dimensional and obey the Euler equations of fluid dynamics on a flat bounded domain.
Hence, the Coriolis force it is often taken in the $\beta$-plane approximation.
In such a way the fluid equations are localized at some latitude on the planet.
However, in the $\beta$-plane approximation some global effects and symmetries of the spherical geometry are clearly lost.
For example, the algebraic structure of the Rossby-Haurwitz waves cannot be taken into account \cite{Hau1940}.

Since the work of Arnold \cite{arn}, it is known that the Euler equations of fluid dynamics have a precise characterization in terms of infinite dimensional geodesic flow.
The 2D Euler equations, in the equivalent form in terms of vorticity can be written as:
\begin{equation}\label{eq:Euler_eqn}
\begin{array}{ll}
&\dot{\omega}=\lbrace\psi,\omega\rbrace\\
&\Delta\psi=\omega,
\end{array}
\end{equation}
where $\omega$ is the vorticity\footnote{In the whole paper, when we write vorticity we understand absolute vorticity.}, $\psi$ the stream function and $\lbrace\cdot,\cdot\rbrace$, $\Delta$ are respectively the Poisson bracket and the Laplace-Beltrami operator on a sphere.
Fundamental in the phenomenology of a 2D fluid are the conservation laws of equations \eqref{eq:Euler_eqn}.
Those can be understood within the theory of symplectic reduction \cite{ArKh1998},\cite{MarWei1983}.
In the 90s and early 2000s  \cite{Zei1991, Zei2004}, exploiting the theory of geometric quantization, Vladimir Zeitlin managed to derive a finite dimensional analogue of the equations \eqref{eq:Euler_eqn}.
His model has been successfully solved numerically, providing important insight into the statistically relevant quantities of the Euler equations \cite{AbMa2003},\cite{MoVi2020}.

In this paper, we present a purely algebraic theory for the spontaneous appearance of zonal jets on a spherical domain.
Starting with the Zeitlin model, we are able to characterize the different jets' structures in terms of decomposition of reductive Lie algebras.
Our results show that the mechanism for the zonal jets' formation relies on some specific perturbation or interaction of Rossby-Haurwitz waves, which causes a turbulent regime driven by the \textit{inverse energy cascade} \cite{Fjo1953, Kr1967}.
The inverse cascade is a peculiar phenomenon of 2D fluids which depends on the conservation of energy and enstrophy (the $L^2$ norm of the vorticity).
It is responsible for the accumulation of energy at the lowest possible frequencies (or equivalently at the largest possible spatial scale).
Surprisingly, the rotation of the sphere is not necessary to obtain the fluid evolving into a zonal state.
Since our results are purely algebraic, it is straightforward to extend them from the Zeitlin model to the original Euler equations (see Section~\ref{sec:EE_zonal_flows}).
The core of our theory can be described looking at the commutation rules of spherical harmonics $Y_{lm}$, in which the vorticity field is decomposed:
\[
\omega = \sum_{l=1}^\infty\sum_{m=-l}^l \omega^{lm}Y_{lm}.
\]
Indeed, in the usual wave number notation $(l,m)$, for $l=1,2,\ldots$ and $m=-l,\ldots,l$, we have that the commutation rules of spherical harmonics are such that the Poisson bracket of $Y_{l,m}$ and $Y_{l',m'}$ has non-zero Fourier coefficients only for harmonics $Y_{l'',m''}$, with wave numbers of the form $m'' = m+m'$ and $l''\equiv l+l'+1 \mbox{ mod }2$ \cite{RiSt2014}.
This simple fact implies that the interaction of Rossby-Haurwitz waves (which are combinations of harmonics with the same wavenumber $l$ and possibly harmonics with wavenumber $l=1$) may be restricted to some specific part of the phase space.
In the Zeitlin model, this reduced phase space corresponds to some space of "striped" matrices.
Since the main diagonal represents the zonal component of the discrete vorticity, the further the stripes are each other, the more the vorticity is likely to have a zonal structure.
Numerical experiments (see Section~\ref{sec:num_sim}) show that in the reduced phase space the fluid tends towards a quasi-zonal state, or equivalently in the Zeitlin model, towards a concentration of the vorticity on the main diagonal.
The vorticty left on the off-diagonals represents intra-zonal vortices, analogous to those observed by Williams \cite{Wil1975}.
This process of quasi-zonalization can be understood both in the continuous and in the discrete model in terms of inverse energy cascade, which tends to accumulate the vorticity at the larger scales.
Indeed, when the discrete vorticity has a striped structure with gap equal to some $k\in\mathbb{N}_0$, the smallest possible positive wavenumber $m$ is equal to $k$.
This implies that for wavenumbers $l<k$ the only possible $m$ is $m=0$, i.e. a zonal vorticity.
Hence, we show in Section~\ref{sec:Pert_quant_RH_waves} that, for $k>2$, the inverse energy cascade tends to accumulate vorticity into the zonal component at the lowest possible frequencies.
In Section~\ref{sec:EE_zonal_flows}, we show that these considerations can be directly extended to the continuous model.

Our results show that the formation of zonal jets has a deep connection with the algebraic properties of the spherical harmonics.
In literature, similar analysis have already been performed in order to understand the mechanism of resonance Rossby-Haurwitz waves \cite{ObuYam2019,RezPitKar1993} and the formation of zonal flows \cite{BurDeLLyn2008}. 
Furthermore, we show that even for zero-momentum vorticity on a non-rotating sphere, for some solutions of the Euler equations there exists a proxy for the Rhines scale, which depends on the distribution of the energy at the low zonal frequencies.
%In conclusion, the onset and the Rhines scale of zonal jets is due to the rotation of a planet, as the rotation determines a specific direction in which the Rossby-Haurwitz waves are perturbed.
%At the same time, the mechanism of zonalization of the flow is mainly determined by the algebraic properties of the spherical harmonics and the conservation laws of the 2D fluid dynamics, responsible of the inverse energy cascade.

%The title of your section 2
\section{2D Euler equations}\label{sec:quant_2D_Euler}
In this section we introduce the fluid model that we consider in the whole paper.
We assume that our fluid is bi-dimensional, homogeneous, incompressible and inviscid.
Even if none of this assumptions is attained by a real fluid, we take this very ideal model in order to capture the core of its dynamical properties. 
Nevertheless, 2D models are very useful in the study of barotropic atmospheric and oceanic flows.
Furthermore, in this work we are not interested in the role played by various sources or sinks of energy and we only focus our attention on the kinematic properties of the fluid.
Under this assumptions, the evolution of the fluid can be described by the Euler equations on a rotating sphere $\Ss^2$.
We regard $\Ss^2$ as embedded in the Euclidean space $\Rr^3$, where we have fixed a Cartesian orthonormal frame $\lbrace\textbf{e}_x,\textbf{e}_y,\textbf{e}_z\rbrace$.
The Euler equations in the vorticity formulation are:
\begin{equation}\label{eq:Euler_eqn2}
\begin{array}{ll}
&\dot{\omega}=\lbrace\psi,\omega\rbrace\\
&\Delta\psi=\omega - f,
\end{array}
\end{equation}
where $\omega\in C^1((0,\infty),C_0^\infty(\Ss^2))$ is the absolute vorticity, $\psi$ is the unique solution in $C_0^\infty(\Ss^2)$ of the second equation (here the $0$ denote the zero-mean of the function) called the stream function and $f(p)=2\Omega \textbf{e}_z\cdot p$ is the Coriolis vorticity.
The bracket is explicitly given by:
\[
\lbrace\psi,\omega\rbrace_p = p\cdot(\nabla\psi\times\nabla\omega),
\]
for any $p\in\Ss^2$, and with the gradient taken in $\Rr^3$, extending constantly on rays the functions defined on $\Ss^2$.
Equations \eqref{eq:Euler_eqn2} posses the following integrals of motion:
\[ H = -\dfrac{1}{2}\int_{\Ss^2}\psi(\omega-f) dS \hspace{1cm} \mathbf M = \int_{\Ss^2} \omega\mathbf{e}_z dS \hspace{1cm} C_k(\omega) = \int_{\Ss^2}\omega^k dS,
\]
for $k=1,2,\ldots$.
$H$ is the Hamiltonian (or kinetic energy), $\textbf M$ is the angular momentum\footnote{If $f=0$ all the three projections of $\omega$ onto $\textbf{e}_x,\textbf{e}_y,\textbf{e}_z$ are conserved.}, $C_k$ are the Casimir functions.
In order to perform our analysis, we introduce the $L^2$ orthonormal basis of $C_0^\infty(\Ss^2)$, given by the spherical harmonics:
\[
Y_{lm}=\sqrt{\dfrac{2l+1}{4\pi}\dfrac{(l-m)!}{(l+m)!}}P^m_l(\cos\theta)e^{im\phi},
\]
where $P^m_l$ are the associate Legendre polynomials, for $l\geq 1$ and $m=-l,\dots,l$.
The coordinates $\theta$ and $\phi$ are the usual inclination and azimuthal angles.
Then, the vorticity $\omega$ can be decomposed as:
\[
\omega = \sum_{l=1}^\infty\sum_{m=-l}^l \omega^{lm}Y_{lm}.
\]
It was first worked out by Jens Hoppe in his PhD thesis \cite{hopPhD}, that the Poisson bracket on the sphere defined above admits a finite dimensional approximation in terms of matrix Lie algebras.
For any $N\geq 1$, let us consider the $N\times N$ complex matrices with zero trace. 
Then, we can define a projection $p_N:C_0^\infty(\Ss^2,\Cc))\rightarrow\SL(N,\Cc)$, defined by 
$p_N:Y_{lm}\mapsto \mbox{i}T^N_{lm}$, where:
\[
(T^N_{lm})_{m_1m_2}=(-1)^{N/2-m_1}\sqrt{2l+1} 
\left( \begin{array}{ccc}
N/2 & l & N/2 \\
-m_1 & m & m_2 
\end{array} \right),
\]
for every $l=1,\ldots,N-1$ and $m=-l,\ldots, l$, and the round bracket is the Wigner 3j-symbol. 
For $l\geq N$, $p_N(Y_{lm})=0$.
The $T^N_{lm}$ form an orthonormal basis of $\SL(N,\Cc)$ with respect to the Frobenius inner product.
Furthermore, if we restrict the projections $p_N$ to $C_0^\infty(\Ss^2,\Rr)$, then the image of $p_N$ is $\SU(N)$, i.e. the anti-hermitian matrices with zero trace.
Finally, we have that \cite{BoMeSc1994}:
\[
\lim_{N\rightarrow\infty}||p_N\lbrace f,g\rbrace - [p_N f,p_N g]_N || \rightarrow 0,
\]
for every $f,g\in C_0^\infty(\Ss^2,\Rr)$, where $[\cdot,\cdot]_N = N^{3/2}[\cdot,\cdot]$ and the norm is the spectral matrix norm. 
Using this formalism, Vladimir Zeitlin introduced his discrete model for the Euler equations \cite{Zei2004}, for any $N\in\Nn$: 
\begin{equation}\label{eq:Euler_quant_Cor}
\dot{W}=[\Delta^{-1}_N (W-F),W]_N,
\end{equation}
where $W\in C^1((0,\infty),\SU(N))$, $F=2i\Omega T_{10}$ and $\Delta^{-1}_N$ is the inverse of a suitable discrete Laplacian defined with the same spectral properties of the spherical one:
\[\Delta^{-1}_NT^N_{lm}=-1/(l(l+1))T^N_{lm},\] 
for any $l=1,...,N$, $m=-l,...,l$.
Equations \eqref{eq:Euler_quant_Cor} posses the following integrals of motion, analogous to those of the equations \eqref{eq:Euler_eqn2}:
\[ H(W)=\frac{1}{2}\Tr(\Delta^{-1}_N (W-F) (W-F)^\dagger). \hspace{1cm} \textbf{M} = W_{10}T_{10} \hspace{1cm} F_k(W)=\Tr(W^k),
\]
for $k=2,\ldots N$. Up to a normalization constant depending on $N$, these quantities converge to the powers of the continuous vorticity (see \cite{CrMiWe1992}).\footnote{If $F=0$ the three components of $W$, $\left( W_{1-1}, W_{10}, W_{11}\right)$ are conserved.}

\subsection{The dynamical irrelevance of the Coriolis force}\label{subsec:Coriolis_force}
In this paragraph, we show that for both the Euler equations \eqref{eq:Euler_eqn2} and their quantized version \eqref{eq:Euler_quant_Cor} we can always assume a zero Coriolis force.
This is not the case of the $\beta$-plane approximation, where the rotational equivariance around the $\textbf{e}_z$-axis of equations \eqref{eq:Euler_eqn2} is broken.
Indeed, this SO(2) symmetry of equations \eqref{eq:Euler_eqn2} and \eqref{eq:Euler_quant_Cor} allows a simple time-dependent change of coordinates which takes away the Coriolis term, while keeping invariant the initial condition.
Consider the rotating Euler equations \eqref{eq:Euler_eqn2}. They can be written as:
\begin{equation*}
\begin{array}{ll}
\partial_t \omega &= \lbrace\Delta^{-1}(\omega-f),\omega\rbrace\\
&=\lbrace\Delta^{-1}\omega,\omega\rbrace + \lbrace\frac{1}{2}f,\omega\rbrace
\end{array}
\end{equation*}
Let $\phi_{-\frac{1}{2}ft}$ be the diffeomorphism generated by Hamiltonian vector field $X_{-\frac{1}{2}f}$, which corresponds to a rotation with respect to the axis $\mathbf{e}_z$ axis of an angle $\Omega t$. 
The RHS of the latter equation is equivariant w.r.t. the coadjoint action $\phi_{-\frac{1}{2}ft}$, for any $t\geq 0$. 
Hence, applying $\phi^*_{-\frac{1}{2}ft}$, we obtain:
\begin{equation*}
\begin{array}{ll}
\phi^*_{-\frac{1}{2}ft}\partial_t \omega &= \phi^*_{-\frac{1}{2}ft}(\lbrace\Delta^{-1}\omega,\omega\rbrace + \lbrace\frac{1}{2}f,\omega\rbrace)\\
&=\lbrace\Delta^{-1}\widetilde{\omega},\widetilde{\omega}\rbrace + \lbrace\frac{1}{2}f,\widetilde{\omega}\rbrace,
\end{array}
\end{equation*}
where $\widetilde{\omega}=\phi^*_{-\frac{1}{2}ft}\omega$. 
From the identity:
\[\partial_t\phi^*_{-\frac{1}{2}ft}\omega = \lbrace\frac{1}{2}f,\widetilde{\omega}\rbrace + \phi^*_{-\frac{1}{2}ft}\partial_t\omega,\]
we conclude that $\widetilde{\omega}$ satisfies the following non-rotating Euler equations:
\begin{equation}\label{eq:non_rot_EE}
\begin{array}{ll}
\partial_t \widetilde{\omega}&=\lbrace\Delta^{-1}\widetilde{\omega},\widetilde{\omega}\rbrace\\\widetilde{\omega}(0)&=\omega(0).
\end{array}
\end{equation}
Hence, we have proved the following proposition:
\begin{proposition}
Let $\omega\in C^1([0,\infty),C_0^\infty(\mathbb{S}^2))$ be a solution of \eqref{eq:Euler_eqn2}, for $f=2\Omega \cos\theta$, $\Omega>0$. Then $\widetilde{\omega}:=\phi^*_{-\frac{1}{2}ft}\omega=\omega(\phi -\Omega t,\theta)$ is a solution of \eqref{eq:non_rot_EE}.
\end{proposition}
Let us now consider equations \eqref{eq:Euler_quant_Cor}.
The RHS of the equations \eqref{eq:Euler_quant_Cor} is equivariant w.r.t. the coadjoint action of $\exp(-\frac{1}{2}Ft)$, for any $t\geq 0$.
Hence, multiplying both sides by $\exp(\frac{1}{2}Ft)(\cdot)\exp(-\frac{1}{2}Ft)$, we obtain:
\begin{equation*}
\begin{array}{ll}
\exp(-\frac{1}{2}Ft)\partial_t W\exp(\frac{1}{2}Ft) &= \exp(-\frac{1}{2}Ft)[\Delta^{-1}W,W]_N\exp(\frac{1}{2}Ft)+\\
&+\exp(-\frac{1}{2}Ft)[\frac{1}{2}F,W]_N\exp(\frac{1}{2}Ft)\\
&=[\Delta^{-1}\widetilde{W},\widetilde{W}]_N+[\frac{1}{2}F,\widetilde{W}]_N,
\end{array}
\end{equation*}
where $\widetilde{W}=\exp(-\frac{1}{2}Ft)W\exp(\frac{1}{2}Ft)$. From the identity:
\[\partial_t\widetilde{W} = -\frac{1}{2}F\widetilde{W} + \exp(-\frac{1}{2}Ft)\partial_t W\exp(\frac{1}{2}Ft) + \widetilde{W}\frac{1}{2}F ,\]
we conclude that $\widetilde{W}$ satisfies the following quantized non-rotating Euler equations:
\begin{equation}
\begin{array}{ll}\label{eq:Euler_quant_no_Cor}
\partial_t \widetilde{W}&=[\Delta^{-1}\widetilde{W},\widetilde{W}]_N\\
\widetilde{W}(0)&=W(0).
\end{array}
\end{equation}

Hence, we have proved the following proposition:
\begin{proposition}\label{prop:qunat_rot_nrot}
Let $W\in C^1([0,\infty),\SU(N))$ be a solution of \eqref{eq:Euler_quant_Cor}, for $F=2\Omega i T_{10}$, $\Omega>0$. Then $\widetilde{W}:=\exp(-\frac{1}{2}Ft)W\exp(\frac{1}{2}Ft)$ is a solution of \eqref{eq:Euler_quant_no_Cor}.
\end{proposition}

Therefore, all the phenomena happening in the non-rotating Euler equations can be found in the rotating ones and viceversa.
In particular, a vorticity field with zero angular momentum on a rotating field can be simply studied as zero angular momentum vorticity field on a non-rotating sphere.

\subsection{Zonal and quasi-zonal flows}\label{subsec:Zonal_quasi_zonal_flows}
In this paragraph we define the zonal and quasi-zonal flows as solutions of the Euler equations \eqref{eq:Euler_eqn2}, and their quantized counterpart as solutions of the equations \eqref{eq:Euler_quant_Cor}.
\begin{definition}\label{def:zf_qzf}
A solution $\omega$ to the Euler equations \eqref{eq:Euler_eqn2} is said to be a zonal flow if it is invariant under rotations about the axis $\textbf{e}_z$.
A solution $\omega$ to the Euler equations \eqref{eq:Euler_eqn2} is said to be a quasi-zonal flow if the complement of its zonal part has smaller energy norm than its zonal part.
\end{definition}
It is clear that the zonal flows are steady solutions of the Euler equations \eqref{eq:Euler_eqn2}.
The definition of quasi-zonal flow wants to emphasize that most of the energy is concentrated in the zonal part of the fluid and hence its behaviour is expected to be "quasi-zonal".
An important class of quasi-zonal flows is given by Rossby-Haurwitz waves with strong angular momentum (see next paragraph).
In Section~\ref{sec:EE_zonal_flows}, an other class of solutions evolving into quasi-zonal flows is presented.
In the quantized case, we can give the following definitions:
\begin{definition}\label{def:quant_zf_qzf}
A solution $W$ to the quantized Euler equations \eqref{eq:Euler_quant_Cor} is said to be a quantized zonal flow if it is a diagonal matrix in the $T_{lm}$ basis.
A solution $W$ to the Euler equations \eqref{eq:Euler_quant_Cor} is said to be a quantized quasi-zonal flow if the complement of its zonal part has smaller energy norm than its zonal part.
\end{definition}
Analogously to the continuous case, the quantized zonal flows are steady solutions of the equations \eqref{eq:Euler_quant_Cor} and a class of quasi-zonal flows is given by quantized Rossby-Haurwitz waves with strong angular momentum (see next paragraph).
In Section~\ref{sec:sliced_matrix_subalgebras_and_zonal_flows}, an other class of solutions evolving into quasi-zonal flows is presented and classified in terms of reductive Lie algebras.
The definitions~\ref{def:zf_qzf} and \ref{def:quant_zf_qzf} are consistent, in the sense that for $N\rightarrow\infty$ the quantized zonal and quasi-zonal flows converge to the continuous ones.
In fact, the diagonal matrices approximate the zonal flows and since the discrete energy coincides up to truncation with the continuous one, for $N$ sufficiently large, the approximating sequence of a quasi-zonal flow becomes a sequence of quantized quasi-zonal flows.

\subsection{Rossby-Haurwitz waves}\label{subsec:Rossby_waves}
Rossby-Haurwitz waves are exact solutions to the Euler equations \eqref{eq:Euler_eqn2}. 
They are defined in terms of spherical harmonics as \cite{Hau1940}:
\begin{equation}\label{RH_waves}
	\omega(\phi,\theta,t) = C f + \sum_{m=-l}^l \omega^{lm}Y_{lm}(\phi+2\Omega\alpha_l t,\theta)
\end{equation}
where $\alpha_l=\frac{1}{2}\left(\frac{2C}{l(l+1)}-C+1\right)$, $\omega^{lm}\in\Cc$, $C\in\Rr$ and $l=1,2,\dots$.
The fact that \eqref{RH_waves} are exact solutions to \eqref{eq:Euler_eqn2} depends only on the algebraic properties of the Poisson bracket of the spherical harmonics. 
Indeed, it is not hard to check that we get an analogous class of exact solutions to \eqref{eq:Euler_quant_Cor} in terms of the spherical matrix basis $T^N_{lm}$:
\begin{equation}\label{RH_waves_quant}
W(t) = C\cdot F + \exp(-\alpha_l N^{3/2} F\cdot t)\sum_{m=-l}^l W^{lm}\mathrm{i} T^N_{lm}\exp(\alpha_l N^{3/2} F\cdot t)
\end{equation}
where $\alpha_l=\frac{1}{2}\left(\frac{2C}{l(l+1)}-C+1\right)$, $W^{lm}\in\Cc$, $C\in\Rr$ and $l=1,2,\dots,N-1$ and $\exp$ is the usual matrix exponential.
We call these solutions \emph{quantized Rossby-Haurwitz waves}.

In the next sections, we determine the evolution of the Rossby-Haurwitz waves under some specific perturbations.
The interest in such waves is motivated by the studies in the mechanism of resonance of Rossby-Haurwitz waves \cite{ObuYam2019,RezPitKar1993}.
In particular, for Rossby-Haurwitz waves with non-zero coefficients only for some $m$ multiple of a fixed integer $k>2$, we show that a perturbation which retains the vorticity symmetries leads the fluid toward a quasi-zonal flow.

\section{Quasi-zonal flows in the quantized Euler equations}\label{sec:EE_quant_zonal_flows}

\subsection{Striped matrix subalgebras and zonal jets}\label{sec:sliced_matrix_subalgebras_and_zonal_flows}

In this section, we analyse a particular subspace of solutions of the equations \eqref{eq:Euler_quant_no_Cor}, which plays a central role in the study of quantized quasi-zonal flows.
In particular, we show that equations \eqref{eq:Euler_quant_no_Cor} have a reduced dynamics into some Lie subalgebra of $\SU(N)$ of striped (or banded) matrices. 
Since these subalgebras are reductive, we can perform a complete classification of them.

Let us consider equations \ref{eq:Euler_quant_no_Cor}.
Since $W\in\SU(N)$, we have the further symmetry of the coefficients: $W^{l-m}=(-1)^m\overline{W}^{lm}$. 
This symmetry comes from the fact that $(iT^N_{lm})^\dagger=(-1)^miT^N_{l-m}$.
Hence, we can consider the respective basis elements in $\SU(N)$ defined as:
\[
R^N_{lm}=\left\{\begin{array}{lll}
\dfrac{1}{\sqrt{2}}(T^N_{lm}-(-1)^mT^N_{l-m}) \hspace{.5cm}\mbox{ for } m>0\\
iT^N_{l0} \hspace{3.4cm}\mbox{ for } m=0\\
\dfrac{i}{\sqrt{2}}(T^N_{l-m}+(-1)^mT^N_{lm}) \hspace{.5cm}\mbox{ for } m<0\\
\end{array}
\right.
\]
for any $l=1,\ldots N-1,m=-l,\dots,l$.
The matrices $R^N_{lm}$ have a banded structure, having non-zero entries only in the $\pm m$~diagonals.

Let $\Dd(N,k)$ be the subalgebra of $\UU(N)$, for some $N>0$ and $0<k\leq N$, defined as follows. 
Let $\Dd(N,k)$ be the set of all the matrices in $\UU(N)$ (seen as the $N\times N$ skew-Hermitian matrices) that have non-zero entries only in the $0^{th},\pm k^{th},\pm 2k^{th},\ldots$ diagonals. It is straightforward to check that the bracket of two matrices of this kind remain of the same form. Therefore $\Dd(N,k)$ is a Lie subalgebra of $\UU(N)$.
Moreover, $\Dd(N,k)$ is a reductive Lie algebra, being closed under complex conjugate transpose \cite[Prop. 6.28]{Kna1996}.
The following theorem gives the reductive structure of $\Dd(N,k)$.
\begin{theorem}\label{thm:DNk}
\[
\Dd(N,k) \cong \UU\left(d\right)^{k-r}\oplus \UU\left(d+1\right)^r\cong \SU\left(d\right)^{k-r}\oplus\SU\left(d+1\right)^r\oplus\UU(1)^k.
\]
where $d=\lfloor\frac{N}{k}\rfloor$ and $r=N-kd$, where $\lfloor\cdot\rfloor$ denotes the integer part. In particular, $\mbox{dim}(\Dd(N,k))=kd^2+2rd + r = d(N+r)+r$.
%If $k\mid n$ then:
%\[
%\GG \cong \UU\left(\frac{n}{k}\right)^{\oplus k}.
%\]
%If $k\nmid n$ then:
%\[
%\GG \cong \oplus_{h=1}^{[n/k]+1} \UU(h)^{l_h}.
%\]
%where $l_h\geq 0$ are such that $n=\sum_{h=1}^{[n/k]+1} hl_h$ and are recursively determined as it follows:
%\begin{enumerate}
%\item If $k\nmid n$, then let $h:=[n/k]+1$, $l_h:=l_h + 1$ and $n:=n-h$;
%\item if $n>0$, then let $h:=[n/k]+1$, $l_h:=l_h + 1$, and $n:=n-h$; else STOP;
%\item repeat Step 2.
%\end{enumerate}
\end{theorem}
\proof
\textbf{Step 1.} The Lie algebra $\Dd(N,k)$ can be decomposed into $k$ subalgebras. Let $A,B\in\Dd(N,k)$ and consider $C=[A,B]\in\Dd(N,k)$. Then, for any $a=1,..,k$: 
\begin{equation}\label{eq:strip_matrix_mult}
C_{a+ik,a+jk}=\sum_{h=0}^{\lfloor\frac{N-a}{k}\rfloor}A_{a+ik,a+hk}B_{a+hk,a+jk}-B_{a+ik,a+hk}A_{a+hk,a+jk},\end{equation}
for all $i,j=0,...,\lfloor\frac{N-a}{k}\rfloor$. Hence, defining 
\[\GG_a=\lbrace M\in\Dd(N,k) \mbox{ s.t. } M_{i,j}\neq 0 \mbox{ iff } i\equiv a \mbox{ mod } k,j\equiv a  \mbox{ mod } k\rbrace,
\]
we see that for any $a=1,...,k$ the $\GG_a$ are closed under the matrix commutator and that $[\GG_a,\GG_b]=0$, for $a\neq b$. In fact, $AB=0$, for any $A\in\GG_a,B\in\GG_b$, for $a\neq b$. Finally, since any element in $\Dd(N,k)$ is a linear combination of elements in the $\GG_a$, for some $a=1,...,k$, we conclude that:
\[
\Dd(N,k) = \oplus_{a=1}^k \GG_a.
\]
\begin{figure}[h!]
\begin{minipage}{1\textwidth}
\[\Dd(7,3)=\left\{A=\left[\begin{matrix}
	  * & 0 & 0 & * & 0 & 0 & *\\
      0 & x & 0 & 0 & x & 0 & 0\\
      0 & 0 & = & 0 & 0 & = & 0\\
      * & 0 & 0 & * & 0 & 0 & *\\
      0 & x & 0 & 0 & x & 0 & 0\\
      0 & 0 & = & 0 & 0 & = & 0\\
      * & 0 & 0 & * & 0 & 0 & *
\end{matrix}
\right], A\in\UU(7)\right\}\cong \UU\left(2\right)^{2}\oplus \UU\left(3\right)
\]
\end{minipage}
\caption{Example of a decomposition of $\Dd(7,3)$ as explained in Step~1, where the symbols $\lbrace*,x,=\rbrace$ denote the only possible non-zero entries.}
\end{figure}
\noindent\textbf{Step 2.} It is clear that $\GG_a$ are closed under complex conjugate transpose, and no further restriction is present. Therefore, $\GG_a\cong\UU(n_a)$, for some $0<n_a\leq N$. \\
\textbf{Step 3.} It is straightforward to check that $n_a=\lfloor\frac{N-a}{k}\rfloor+1$, for any $a=1,...,k$. Therefore, defining $r=N-k\lfloor\frac{N}{k}\rfloor$, we have that for $a=1,...,r$, $n_a=\lfloor\frac{N}{k}\rfloor+1$ and for $a=r+1,...,k$, $n_a=\lfloor\frac{N}{k}\rfloor$.
%The procedure goes as follows. 
%Define $\GG_1<\GG$ such that $A\in\GG_1$ iff $A_{ij}\neq 0$ iff $i,j=1+k,1+2k,\ldots$. 
%If $k\mid n$, $\GG_1\cong\UU(n/k)$ and if $k\nmid n$, $\GG_1\cong\UU([n/k]+1)$. 
%Then, define $\GG_2<(\GG_1^\perp\cap\GG)$, such that $A\in\GG_2$ iff $A_{ij}\neq 0$ iff $i,j=2+k,2+2k,\ldots$. 
%If $k\mid n-1$, $\GG_1\cong\UU((n-1)/k)$ and if $k\nmid n-1$, $\GG_1\cong\UU([(n-1)/k]+1)$. 
%Iterate the procedure until $\GG_m=\lbrace 0\rbrace$. 
%Clearly the procedure terminates, being $\mbox{dim}((\cup_{i=1}^{m}\GG_i)^\perp\cap\GG)<\mbox{dim}((\cup_{i=1}^{m-1}\GG_i)^\perp\cap\GG)$, for every $m\geq 1$.
\endproof
Let us now recall the abstract definition of the discrete Laplacian $\Delta_N$ and show that it can be restricted to an operator on $\Dd(N,k)$, for any $k=1,\ldots, N$. 
Let $\mathcal{B}=\lbrace X_1,X_2,X_3\rbrace$ be a basis of $\SU(2)$ and let $B$ a non-degenerate, $\mbox{ad}$-inviariant, bilinear form $B$ on $\SU(2)$ (e.g. $B$ can be assumed to be the Killing form).
Then let $\mathcal{B}^\vee=\lbrace X^1,X^2,X^3\rbrace$ be the dual basis of $\mathcal{B}$ with respect to $B$. 
Then, the Casimir element $C$ of $\SU(2)$ is defined as an element of the universal enveloping algebra $\mathfrak{U}(\SU(2))$ of $\SU(2)$ and can be written as 
\[C=\sum_{i=1}^3 X_iX^i.\] 
Notice that this definition of $C$ does not depend neither on the choice of $\mathcal{B}$ nor $B$. 
Consider the irreducible representation $\varrho$ of $\SU(2)$ on $\Cc^n$ and the adjoint representation of $\SU(N)$ on itself. 
Define the discrete Laplacian as:
\[\Delta_N:=\mbox{ad}_{\varrho(C)}:=\sum_{i=1}^3\mbox{ad}_{\varrho(X_i)}\circ\mbox{ad}_{\varrho(X_i)}:\SU(N)\rightarrow\SU(N).\]
Finally, in this notation, the discrete Laplacian $\Delta_N$ acts on $\SU(N)$ as:  
\[\Delta_N(Y)=\mbox{ad}_{\varrho(C)}(Y)=[\varrho(X_1),[\varrho(X_1),Y]] + [\varrho(X_2),[\varrho(X_2),Y]] + [\varrho(X_3),[\varrho(X_3),Y]],\]
for any $Y\in\SU(N)$. 
\begin{theorem}\label{thm:lapl_end}
The discrete Laplacian $\Delta_N$ is a vector space endomorphism of $\Dd(N,k)$.
\end{theorem}
\proof It is clear that:
\[\Dd(N,k)=\mbox{Span}\left\lbrace R^N_{l\pm \alpha k}, \mbox{ for } l=1,\ldots,N,\alpha=0,\ldots,\left\lfloor\frac{l}{k}\right\rfloor\right\rbrace\oplus i\Rr Id.\]
In \cite{HopYau1998} it is shown that $\lbrace R^N_{l\pm \alpha k},Id\rbrace$ are eigenvectors of $\Delta_N$.
Hence, $\Dd(N,k)$ is invariant with respect to the action of $\Delta_N$.
\endproof
\begin{corollary}
The discrete Laplacian $\Delta_N$ is a vector space automorphism of $\Dd(N,k)\cap\SU(N)$. Hence, the quantized Euler equations \eqref{eq:Euler_quant_no_Cor} can be restricted to $\Dd(N,k)\cap\SU(N)$.
\end{corollary}

Since the vorticity has zero mean, we will restrict to $\Dd_0(N,k):=\Dd(N,k)\cap\SU(N)$.

\subsection{Split equations}
Consider the splitting of Theorem~\ref{thm:DNk}, for some $\Dd(N,k)$.
We have the natural projections $\Pi_{d,a},\Pi_{d+1,b}$ respectively onto the subspaces of $\UU(N)$ isomorphic to $\UU(d)$ and $\UU(d+1)$, for $a=1,\ldots,k-r$ and $b=1,\ldots,r$.
From \eqref{eq:strip_matrix_mult}, it is clear that the dynamics of the factors $W_{d,a}=\Pi_{d,a}W, W_{d+1,b}=\Pi_{d+1,b}W$ can be derived from the equations \eqref{eq:Euler_quant_no_Cor} as:
\begin{equation}\label{eq:factor_dynamics}
\begin{array}{ll}
\dot{W}_{d,a} &= [\Delta^{-1}W,W_{d,a}]_N\\
\dot{W}_{d+1,b} &= [\Delta^{-1}W,W_{d+1,b}]_N
\end{array}
\end{equation}
for $a=1,\ldots,k-r, b=1,\ldots,r$ and $W=\sum_{a=1}^{k-r}W_{d,a}+\sum_{b=1}^{r}W_{d+1,b}$.
Hence, from the equations \eqref{eq:factor_dynamics} we see that the the vorticty factors are transported by the same stream function. 
Moreover, equations \eqref{eq:factor_dynamics} are an isospectral flow on $\mathfrak{s}(\UU(d)^{k-r}\oplus\UU(d+1)^r)$. 
Hence there are $(k-r)d+r(d+1)-1=N-1$ conserved quantities corresponding to the Casimirs of $W$.

\subsection{Perturbation and interaction of quantized Rossby-Haurwitz waves}\label{sec:Pert_quant_RH_waves}
In this paragraph, we show how the results in Section~\ref{sec:sliced_matrix_subalgebras_and_zonal_flows} can be used to study some properties of the quantized Rossby-Haurwitz waves.
Let $W$ be a quantized Rossby-Haurwitz wave in some $\Dd_0(N,k)$.
Given $l<N$, the generic (real) form at $t=0$ is:
\begin{equation}\label{eq:RH_waves_quant_Dnk}
W(0) = C\cdot F + \sum_{h\in\Zz,|hk|\leq l} W^{l\hspace{.1cm}hk} R^N_{l\hspace{.1cm}hk}.
\end{equation}
As shown in Section~\ref{sec:sliced_matrix_subalgebras_and_zonal_flows}, the spaces $\Dd_0(N,k)$ are invariant under the quantized Euler equations \eqref{eq:Euler_quant_no_Cor}. 
Hence, it makes sense studying the evolution of a perturbed quantized Rossby-Haurwitz wave defined as in \eqref{eq:RH_waves_quant_Dnk} in the reduced space $\Dd_0(N,k)$.
Furthermore, in the same framework, we can study the interaction of two Rossby-Haurwitz waves defined as in \eqref{eq:RH_waves_quant_Dnk}. 
The key fact is expressed in the following Lemma:
\begin{lemma}\label{lem:RH_interaction}
The bracket closure of $R^N_{lm},R^N_{l'm'}$ is included in $\Dd_0(N,k)$, for $k=gcd(m,m')$.
More precisely, the bracket closure of $R^N_{lm},R^N_{l'm'}$ is  $\Dd_0(N,k)$ unless both $l,l'$ are odd, for which it consists of the linear combination of eigenmodes $l'',m''$ such that $l+l'+1\equiv l'' (2)$ and $m''=\pm hk$, for $h=0,1,2,\ldots$.
\end{lemma}
\proof
The bracket closure of some collection of square matrices $\mathcal{A}=\lbrace A_i\rbrace_{i\in I}$ is the smallest Lie algebra containing the repeated bracketing of elements in $\mathcal{A}$.
Let $\mathcal{A}=\lbrace A,B\rbrace$, for $A=R^N_{lm},B=R^N_{l'm'}$, and define $C=[A,B]$. Then, for any $i,j=1,..,N$: 
\[C_{i,j}=\sum_{k=1}^N A_{i,k}\delta^k_{i\pm m}\delta^k_{j\pm m'}B_{k,j}-B_{i,k}\delta^k_{i\pm m'}\delta^k_{j\pm m}A_{k,j}.\]
Hence, $C_{i,j}\neq 0$ only if $i=j \pm m' \mp m$ or $i=j \pm m \mp m'$.
This implies that the bracket of two banded matrices is still banded. 
Iterating the bracketing, we notice that the smallest bands gap which is possible to reach is: 
\[k=\min_{a,b\in\Zz}\lbrace |am +bm'|,\mbox{ s.t. }|am +bm'|>0\rbrace,\]
which is equal to the greatest common divisor of $m,m'$.

The second part can be straightforwardly proved using Proposition 3.3.12 \cite{RiSt2014}.
\endproof
Consider the quantized Euler equations \eqref{eq:Euler_quant_no_Cor} and let
\begin{equation}
W(0) = C\cdot F + \sum_{h\in\Zz,|hk|\leq l} W^{l\hspace{.1cm}hk} R^N_{l\hspace{.1cm}hk} + \Upsilon,
\end{equation}
be the initial vorticity, where $k,l<N$ and $\Upsilon\in\Dd_0(N,k')$, for some $k'<N$.
$\Upsilon$ is a perturbation that we assume, without loss of generality, with zero angular momentum.
Let us call $M:=C\cdot F $ and $R:=\sum_{h\in\Zz,|hk|\leq l} W^{l\hspace{.1cm}hk} R^N_{l\hspace{.1cm}hk}$.
Then, the perturbation $\Upsilon$ satisfies:
\[
\dot{\Upsilon}=[\Delta_N^{-1}(\Upsilon+R),(\Upsilon+R)]_N + [\Delta_N^{-1}M,\Upsilon+R]_N + [\Delta_N^{-1}(\Upsilon+R),M]_N.
\]
The second term on the right hand side can be taken away via the same calculations of Section~\ref{subsec:Coriolis_force} and so we can study:
\begin{equation}\label{eq:pert_simpl}
\dot{\widetilde{\Upsilon}}=[\Delta_N^{-1}(\widetilde{\Upsilon}+\widetilde{R}),(\widetilde{\Upsilon}+\widetilde{R})]_N + [\Delta_N^{-1}(\widetilde{\Upsilon}+\widetilde{R}),M]_N,
\end{equation}
where $\widetilde{\Upsilon}+\widetilde{R}=\exp(t/2M)(\Upsilon+R)\exp(-t/2M)$.
By the Theorem~\ref{thm:lapl_end} and Lemma~\ref{lem:RH_interaction}, the second term on the right hand side of \eqref{eq:pert_simpl} is such that if $\widetilde{\Upsilon}+\widetilde{R}\in\Dd_0(N,k'')$, for some $k''<N$, then $[\Delta_N^{-1}(\widetilde{\Upsilon}+\widetilde{R}),M]\in\Dd_0(N,k'')$.
Applying Lemma~\ref{lem:RH_interaction} to the first term on the right hand side of \eqref{eq:pert_simpl}, we get that $[\Delta_N^{-1}(\Upsilon+R),(\Upsilon+R)]_N\in \Dd_0(N,gcd(k,k'))$.
In conclusion, the quantized vorticity $W$ evolves in $\Dd_0(N,gcd(k,k'))$.
In particular if $W(0)$ is the sum of two quantized Rossby-Haurwitz waves of the form \eqref{eq:RH_waves_quant_Dnk}, they evolve in some space $\Dd_0(N,k'')$, accordingly to the Lemma~\ref{lem:RH_interaction}.

We claim that the generic long-time behaviour of a perturbed quantized Rossby-Haurwitz wave in the setting above is such that it leads the vorticity into a quantized quasi-zonal flow.
In Section~\ref{sec:num_sim}, we provide numerical evidences about this behaviour for $M=0$.
\begin{remark}
The long-time behaviour for the case of $M\neq 0$ can be recovered considering that, up to a constant rotation, there is a one-to-one correspondence of the zero momentum steady states with the non-zero momentum steady states.
This can be immediately derived from equation \eqref{eq:pert_simpl}, and observing that in the zero momentum case the steady states are characterized by a functional relation between the vorticity and the stream function.
\end{remark}
The principle which determines the long time evolution of a 2D fluid is called inverse energy cascade \cite{BoVe2012}.
In essence, the non-linearity of the Euler equations pushes the vorticity into the high frequencies, however the conservation of energy and enstrophy prevents that the vorticity is all spread into the high frequencies, and indeed most of the energy is accumulated at the low frequencies.
The same principle can be stated in the quantized case, having conservation of the discrete energy and enstrophy.
More precisely, we can state the following inequalities, analogous to those in \cite[2.3.2]{BoVe2012}.
Let us decompose the energy and the enstrophy accordingly to the wavenumber $l$:
\[
H = \sum_{l=1}^N H(l), \hspace{1cm} E = \sum_{l=1}^N E(l),
\]
where $H(l) = \frac{1}{2l(l+1)}E(l) =  \frac{1}{2l(l+1)}\sum_{m=-l}^l |W^{lm}|^2$.
We define the centroids of $H$ and $E$ as:
\[
c_H = \frac{1}{H}\sum_{l=1}^N lH(l), \hspace{1cm} c_E = \frac{1}{E}\sum_{l=1}^N lE(l).
\]
\[
v_H = \frac{1}{H}\sum_{l=1}^N \frac{1}{l}H(l), \hspace{1cm} v_E = \frac{1}{E}\sum_{l=1}^N \frac{1}{l}E(l).
\]
Then, using the Cauchy-Schwartz inequality, we get:
\begin{equation*}
\begin{array}{ll}
c_H &= \dfrac{1}{H}\sum_{l=1}^N lH(l)= \dfrac{1}{H}\sum_{l=1}^N \sqrt{H(l)}\sqrt{l^2H(l)}\leq \dfrac{1}{H} \sqrt{H}\sqrt{\dfrac{E}{2}}=\sqrt{\dfrac{E}{2H}},\\
c_H &= \dfrac{1}{H}\sum_{l=1}^N \dfrac{E(l)}{2l}=\dfrac{E}{2H}v_E,
\\
1 &= \left(\dfrac{1}{E}\sum_{l=1}^N E(l)\right)^2 = \left(\dfrac{1}{E}\sum_{l=1}^N \dfrac{\sqrt{l}}{\sqrt{l}} E(l)\right)^2\leq c_E v_E,
\\
1 &= \left(\dfrac{1}{H}\sum_{l=1}^N H(l)\right)^2 = \left(\dfrac{1}{H}\sum_{l=1}^N \dfrac{\sqrt{l}}{\sqrt{l}} H(l)\right)^2\leq c_H v_H.
\end{array}
\end{equation*}
From these inequalities we then have:
\begin{equation}\label{eq:cascade_ineq1}
c_H\leq\sqrt{\dfrac{E}{2H}}, \hspace{1cm} c_E\geq\sqrt{\dfrac{E}{2H}},  \hspace{1cm} c_Hc_E\geq\dfrac{E}{2H}.
\end{equation}
\begin{equation}\label{eq:cascade_ineq2}
v_H\geq\sqrt{\dfrac{2H}{E}}, \hspace{1cm} v_E\leq\sqrt{\dfrac{2H}{E}},  \hspace{1cm} v_Hv_E\geq\dfrac{2H}{E}.
\end{equation}
The first inequality of \eqref{eq:cascade_ineq1} puts a constraint on the minimum scale at which the energy can be transferred.
Analogously, the second inequality of \eqref{eq:cascade_ineq1} puts a constraint on the maximum scale at which the enstrophy can be transferred.
The third inequality of \eqref{eq:cascade_ineq1} says that an inverse cascade of the energy must be compensated by a forward cascade of the enstrophy.
Finally, the third inequality of \eqref{eq:cascade_ineq2} says that a forward cascade of the enstrophy must be compensated by  an inverse cascade of the energy.

Hence, from the inequalities \eqref{eq:cascade_ineq1}, \eqref{eq:cascade_ineq2},  we have that in the quantized model the energy accumulates at the low frequencies, while the enstrophy is drifted into the high frequencies due to the non-linearity of the quantized Euler equations. 
In the quantized case, the corresponding basis element of the real frequency $W^{lm}$ is a matrix with non-zero entries only on the $\pm m$ subdiagonal.
Hence, the lower the frequency is, the closer to the main diagonal the support is.
Therefore, the inverse energy cascade pushes the discrete vorticity into a more and more diagonally dominant matrix. 
In particular, in the spaces $\Dd_0(N,k)$, for $k>2$, the larger the $k$ is, the further the first non-zero subdiagonal is from the main diagonal.
Hence, via the inverse energy cascade, the vorticity condensates on the main diagonal.
Since the diagonal matrices correspond to zonal-flows, the larger the $k$ is, the more the flow will be zonal, since for the lowest possible $l$, the only admissible $m$, multiple of $k$, is $m=0$.
Finally, the number $k$ determines the blobs trapped into these zonal bands.
Indeed, the lowest non-zero off-diagonal components correspond to the spherical harmonics $Y_{lk}$, for $l=k,\dots,N$ which have $k$ latitudinal blobs (see Figure~\ref{fig:k=3},\ref{fig:k=7},\ref{fig:k=11},\ref{fig:k=17}).

\section{Quasi-zonal flows in the Euler equations}\label{sec:EE_zonal_flows}
In this section, we extend the results of Section~\ref{sec:sliced_matrix_subalgebras_and_zonal_flows}-\ref{sec:Pert_quant_RH_waves} to the Euler equations \eqref{eq:Euler_eqn2}.
In view of the \textit{weak-uniqueness theorem} of \cite{BoHoScSc1991}, we have that the sequence of $\Dd_0(N,k)$ approximates some subalgebra of $C^\infty_0(\Ss^2)$, for $N\rightarrow\infty$.
More precisely, let $\Dd_0(k)$ be the vector space of of smooth functions, which are linear combinations of spherical harmonics of the form $Y_{l\pm \alpha k}$, for $\alpha=1,2,\ldots$.
Then, we have the following result:
\begin{proposition}\label{prop:Dk_struct}
$\Dd_0(k)$ is a Lie subalgebra of $C^\infty_0(\Ss^2)$, which is approximated by the $\Dd_0(N,k)$, for $N\rightarrow\infty$.
\end{proposition} 
\proof
Let $f\in C^\infty_0(\Ss^2)$. 
Then it is straightforward to check that $f\in\Dd_0(k)$ if and only if, for any $N>0$, $p_N f\in\Dd_0(N,k)$.	
In \cite{BoMeSc1994} it is shown that for every $f,g\in C^\infty_0(\Ss^2)$:
\[
\|p_N\lbrace f,g\rbrace - \left[p_N f,p_N g\right]_N\|\rightarrow 0, \hspace{1cm} \mbox{for } N\rightarrow\infty.
\]
Moreover, for every $f,g\in \Dd_0(k)$, $\left[p_N f,p_N g\right]_N\in\Dd_0(N,k)$, for any $N>0$.
Hence, $\Dd_0(N,k)$ approximates $(\Dd_0(k),\lbrace \cdot,\cdot\rbrace_{|\Dd_0(k)})$, and so by \cite[Prop. 3.3]{BoHoScSc1991} $\Dd_0(k)$ is a Lie subalgebra of $C^\infty_0(\Ss^2)$.
\endproof
\begin{remark}
We notice that Proposition~\ref{prop:Dk_struct} can also be proved directly using the structure constants of $C^\infty_0(\Ss^2)$ in the spherical harmonics basis.
\end{remark}

\subsection{Split equations}
\begin{figure}
\centering
\subfloat[Vorticity field $W_{86,1}(t_{0})$]{
 \includegraphics[width=.5\textwidth]{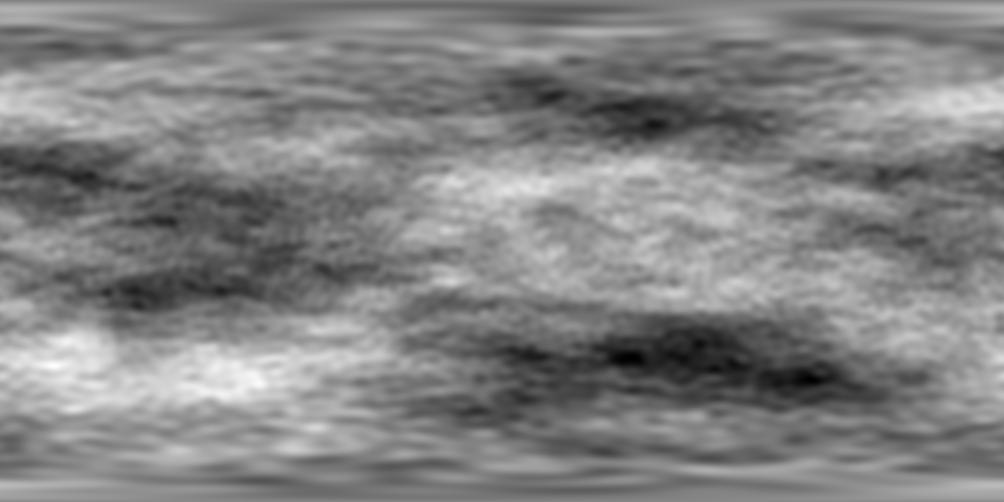}}
\subfloat[Vorticity field $W_{86,1}(t_{end})$]{
  \includegraphics[width=.5\textwidth]{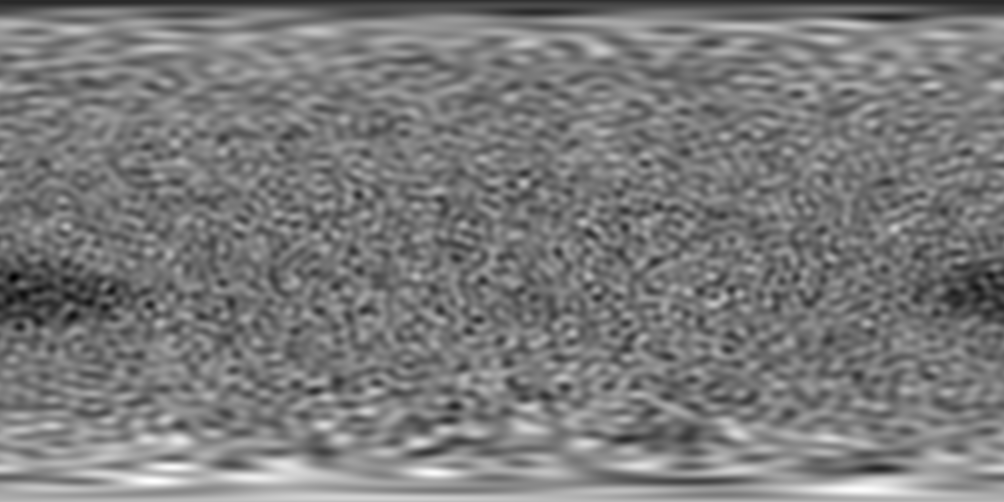}}\\
  \subfloat[Vorticity field $W_{86,2}(t_{0})$]{
  \includegraphics[width=.5\textwidth]{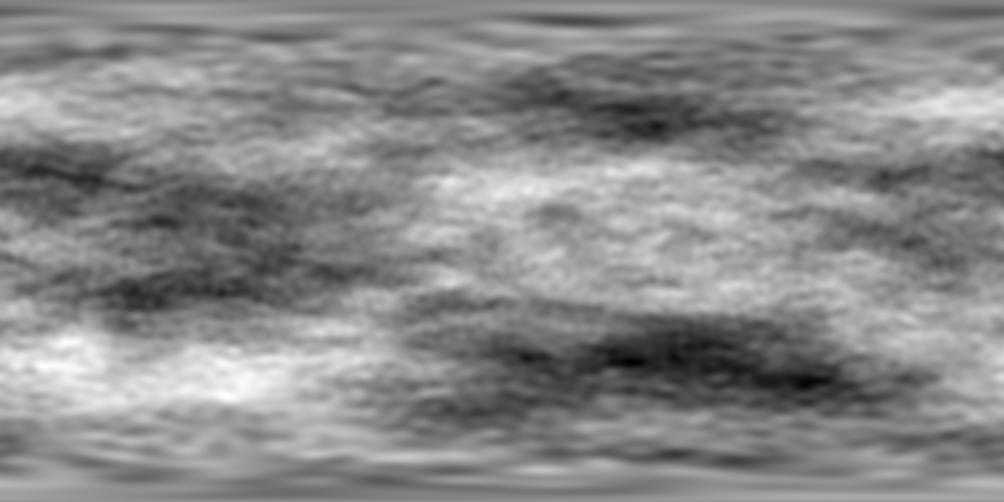}}
\subfloat[Vorticity field $W_{86,2}(t_{end})$]{
  \includegraphics[width=.5\textwidth]{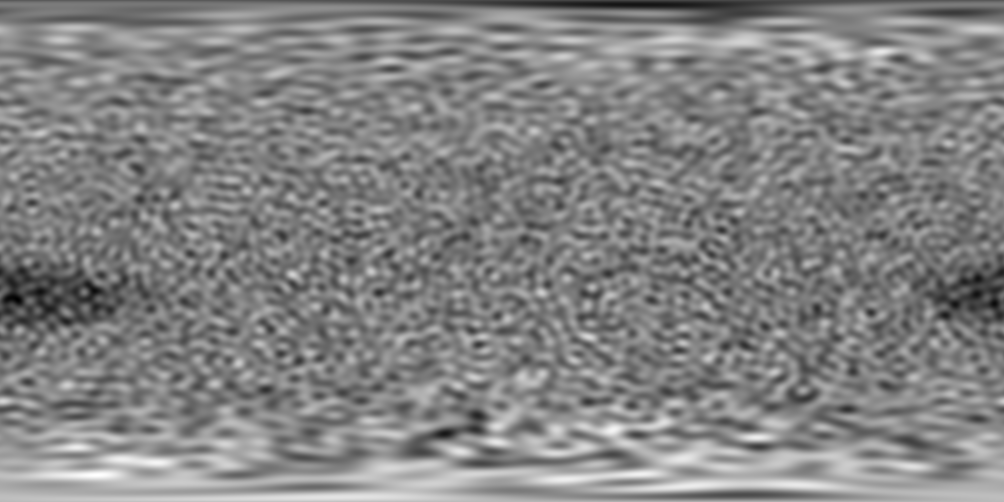}}\\
   \subfloat[Vorticity field $W_{85,1}(t_{0})$]{
  \includegraphics[width=.5\textwidth]{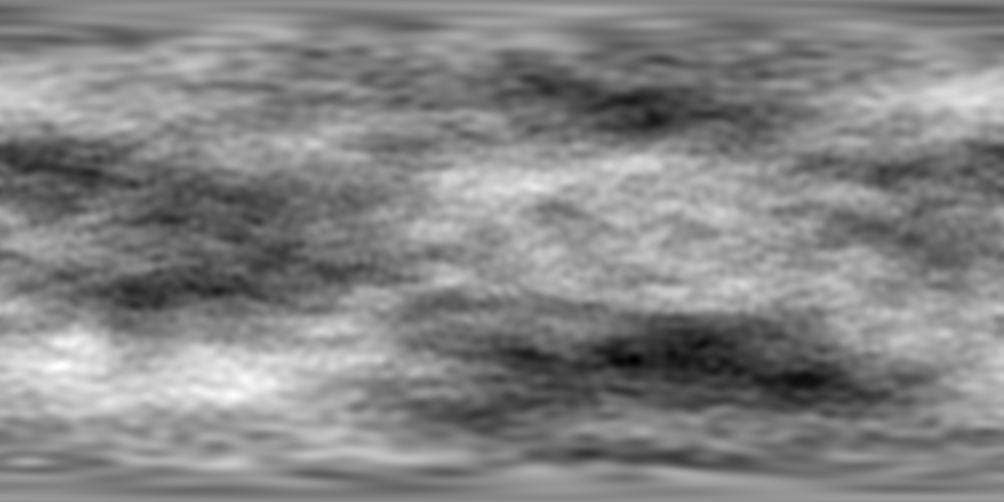}}
\subfloat[Vorticity field $W_{85,1}(t_{end})$]{
  \includegraphics[width=.5\textwidth]{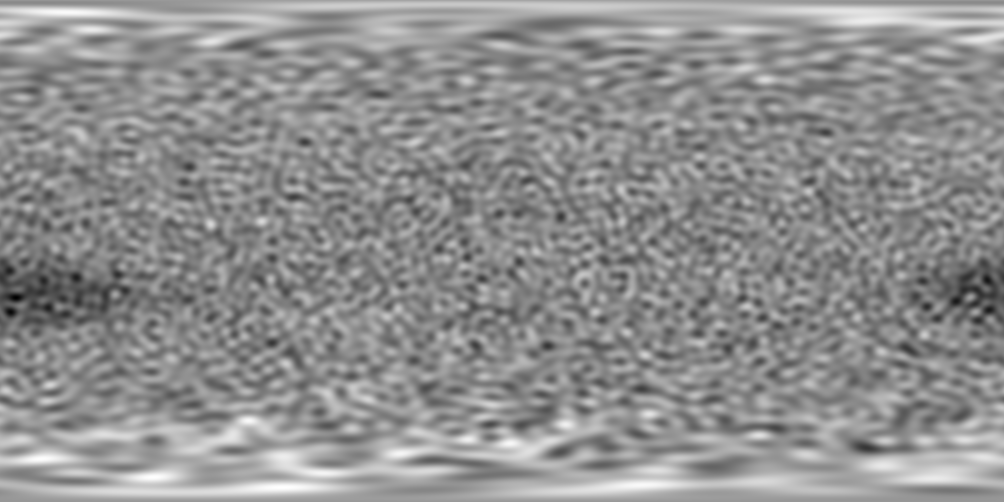}}\\
\caption{Illustration of Theorem~\ref{thm:comp_adj} and Remark~\ref{rem:split_dyn} in $\Dd(257,3)$ for vorticity fields $W_{86,1},W_{86,2},W_{85,1}$ at $t=t_0$ and $t=t_{end}$.}\label{fig:thm5_k=3}
\end{figure}

The result of Theorem~\ref{thm:DNk} has a trivial extension to the infinite dimensional Poisson algebra of $C^\infty_0(\Ss^2)$.
Indeed, we show below that the projections $\Pi_{d,a},\Pi_{d+1,b}$ defined in the previous section, combined with the projections $p_N$, in the limit for $N\rightarrow\infty$, converge almost everywhere to the same limit in $C^\infty_0(\Ss^2)$.
In particular, equations \eqref{eq:factor_dynamics} in the limit for $N\rightarrow\infty$ converge to the same Euler equations.
\begin{lemma}\label{lem:sph_matrix}
Let $T^N_{lm}$ be a basis element in $\SU(N)$ as defined in Section~\ref{sec:quant_2D_Euler}, for some $l=1,\ldots,N-1$ and $m=-l,\ldots,l$. 
Then, for any two adjacent components of some diagonal of $T^N_{lm}$, i.e. for any $a,b,a',b'$ such that $a-b=a'-b'$ and $a'=a\pm 1,b'=b\pm 1$, we have that 
\[
|T^N_{lm,ab}-T^N_{lm,a'b'}|\rightarrow 0,
\]
for $N\rightarrow\infty$.
\end{lemma}
\proof
Let us recall the definition of the (not-normalized) $T^N_{lm}$ as given in \cite{HopYau1998} in terms of the $N\times N$ matrices $X_1,X_2,X_3$, defined as:
\begin{equation}\label{eq:sph_matrix}
\begin{array}{ll}
&(X_1)_{ab}=\dfrac{1}{\sqrt{N^2-1}}\delta_{a}^{b\pm 1}\sqrt{s(s+1)-(-s+b-1)(-s+b-1\pm 1)} \\
&(X_2)_{ab}=\dfrac{\mp i}{\sqrt{N^2-1}}\delta_{a}^{b\pm 1}\sqrt{s(s+1)-(-s+b-1)(-s+b-1\pm 1)} \\
&(X_3)_{ab}=\dfrac{\pm 2}{\sqrt{N^2-1}}\delta_{a}^b(-s+a-1),
\end{array}
\end{equation}
where $s=\frac{N-1}{2}$.
Then,
\begin{equation}\label{eq:matrix_basis_sph_matrix}
T^N_{lm} =\overline{c}_{Nl}\sum_{a_1,\ldots,a_l\in\lbrace 1,2,3\rbrace} c^{(m)}_{a_1\ldots a_l}X_{a_1}\cdot X_{a_2}\cdot\ldots\cdot X_{a_l},
\end{equation}
where the indices $c^{(m)}_{a_1\ldots a_l}$ are the totally symmetric tensor defining the spherical harmonics and $\overline{c}_{Nl}= \sqrt{N}\sqrt{\dfrac{(N^2-1)^l(N-l-1)!}{N+l!}}\rightarrow 1$, for $N\rightarrow\infty$ and every fixed $l$.
It is clear from equation \eqref{eq:sph_matrix} that the difference of any two adjacent components of the same diagonal of $X_1, X_2,X_3$ goes to 0, for $N\rightarrow\infty$.
Hence, from equation \eqref{eq:matrix_basis_sph_matrix} it is straightforward to deduce the same result for any $T^N_{lm}$, for fixed $l,m$.
\endproof
Then, we get the following result.
\begin{theorem}\label{thm:comp_adj}
Let $f\in \Dd(k)$ for some fixed $k\geq 1$.
Then, given any $a,b,a',b'$ such that $a-b=a'-b'=\pm hk$ and $a'=a\pm 1,b'=b\pm 1$, and $h=0,1,\ldots$, we have that: 
\[
|(p_N f)_{ab}-(p_N f)_{a'b'}|\rightarrow 0,
\]
for $N\rightarrow\infty$.
\end{theorem}
\proof
First of all, let us notice that being $f$ differentiable, its Fourier series is absolutely convergent.
Further, the Fourier coefficients $f^{lm}$ determining the components $a,b$ and $a',b'$ such that $a-b=a'-b'=hk$ must have $|m|=hk$.
Hence, fix $\varepsilon>0$ and $a-b=a'-b'=\pm hk$ and $a'=a\pm 1,b'=b\pm 1$.
Then we get:
\begin{equation}
\begin{array}{ll}
|(p_N f)_{ab}-(p_N f)_{a'b'}|&=\sum_{l\geq 1,m=\pm hk}|f^{lm}||T^N_{lm,ab}-T^N_{lm,a'b'}|\\
&=\sum_{1\leq l\leq L,m=\pm hk}|f^{lm}||T^N_{lm,ab}-T^N_{lm,a'b'}|+\\
&\sum_{l>L,m=\pm hk}|f^{lm}||T^N_{lm,ab}-T^N_{lm,a'b'}|\\
&\leq \max\limits_{1\leq l\leq L,m=\pm hk}\lbrace |T^N_{lm,ab}-T^N_{lm,a'b'}|\rbrace\sum_{1\leq l\leq L,m=\pm hk}|f^{lm}|\\
&+\sum_{l>L,m=\pm hk}|f^{lm}|(|T^N_{lm,ab}|+|T^N_{lm,a'b'}|),
\end{array}
\end{equation}
for some $L>0$. 
Since for any $g\in C^\infty_0(\Ss^2)$, $\|p_N g\|\rightarrow\|g\|_\infty$, for $N\rightarrow\infty$ \cite{BoMeSc1994}, the second term can be made smaller than $\varepsilon/2$ for any $L,N$ sufficiently large.
For the first term we can use Lemma~\ref{lem:sph_matrix} and take $N$ possibly larger to make it smaller than $\varepsilon/2$.
\endproof
\begin{remark}\label{rem:split_dyn}
Theorem~\ref{thm:comp_adj} allows us to describe the asymptotic of equations \eqref{eq:factor_dynamics} for $N\rightarrow\infty$.
Indeed, the various factors $W_{d,a}, W_{d+1,b}$ are constructed by projecting $W$ onto the different factors which come from adjacent components, as shown in the proof of Theorem~\ref{thm:DNk}.
Hence, by Theorem~\ref{thm:comp_adj} for $N$ sufficiently large, the factors $W_{d,a}, W_{d+1,b}$ are very close to each other.
In particular, the large scale features which are mainly characterized by small wave numbers $l$ and $m$ are quite similar in any of the $W_{d,a}, W_{d+1,b}$ factors (when seen as elements in $\SU(d),\SU(d+1)$ respectively).
In particular, each of equations~\eqref{eq:factor_dynamics} converges in each Fourier component to the Euler equations \eqref{eq:Euler_eqn}, for $N\rightarrow\infty$ .
Indeed, the various factors differ infinitesimally for $N\rightarrow\infty$, as stated in Theorem~\ref{thm:comp_adj}. 
Hence, the decomposition of the Lie algebras $\Dd(N,k)$ becomes redundant in the limit for $N\rightarrow\infty$ (see Figure~\ref{fig:thm5_k=3}).
Conversely, given possibly very different factors $W_{d,a}, W_{d+1,b}$ as initial data, is equivalent to give a very irregular initial vorticity.
However, numerical observations shows that the inverse energy cascade provides a sort of large scale regularization of the vorticity \cite{MoVi2020}.
Hence, for long times, the different factors tend to look pretty close, at least at the large scales.

In this perspective, the system of equations \eqref{eq:factor_dynamics} can be interpreted as a \textit{Domain decomposition method}, and for very large $N$ an iterative scheme like the Schwarz alternating method could be beneficial. 
This issue is currently under investigation.
\end{remark}

\subsection{Perturbation and interaction of Rossby-Haurwitz waves}\label{subsec:Pert_RH_waves}
In this paragraph, we extend the results of Section~\ref{sec:Pert_quant_RH_waves} to the Euler equations \eqref{eq:Euler_eqn}.
The fundamental facts determining the mechanism of zonalization of the vorticity, i.e. the inverse energy cascade and the algebraic properties of the spherical harmonics, directly extends from $\Dd_0(N,k)$ to $\Dd_0(k)$.

First of all, the inverse energy cascade for the Euler equations \eqref{eq:Euler_eqn} can be derived taking the limit for $N\rightarrow\infty$ for both the inequalities \eqref{eq:cascade_ineq1}, \eqref{eq:cascade_ineq2} (see \cite{BoVe2012}).
Then, let $\omega$ be a Rossby-Haurwitz wave in some $\Dd_0(k)$.
Given $l=1,2,\ldots$, the generic form at $t=0$ is:
\begin{equation}\label{eq:RH_waves_Dnk}
	\omega(\phi,\theta,t) = C f + \sum_{h\in\Zz,|hk|\leq l} \omega^{l\hspace{.1cm}hk}Y_{l\hspace{.1cm}hk}
\end{equation}
As shown above, the spaces $\Dd_0(k)$ are invariant under the Euler equations \eqref{eq:Euler_eqn}. 
Hence, it makes sense studying the evolution of a perturbed Rossby-Haurwitz wave defined as in \eqref{eq:RH_waves_Dnk} in the reduced space $\Dd_0(k)$.
Furthermore, extending the result in Lemma~\ref{lem:RH_interaction} for $N\rightarrow\infty$, we can study the interaction of two Rossby-Haurwitz waves defined as in \eqref{eq:RH_waves_quant_Dnk}. 
\begin{lemma}\label{lem:RH_interaction2}
The bracket closure of $Y_{lm},Y_{l'm'}$ is included in the $\Dd_0(k)$, for $k=gcd(m,m')$.
More precisely, the bracket closure of $Y_{lm},Y_{l'm'}$ is $\Dd_0(k)$ unless both $l,l'$ are odd, for which it consists of the linear combination of eigenmodes $l'',m''$ such that $l+l'+1\equiv l'' (2)$ and $m''=\pm hk$, for $h=0,1,2,\ldots$.
\end{lemma}
\proof See \cite{RiSt2014}.
\endproof

Consider the quantized Euler equations \eqref{eq:Euler_eqn} and let
\begin{equation}
\omega(\phi,\theta,t) = C f + \sum_{h\in\Zz,|hk|\leq l} \omega^{l\hspace{.1cm}hk}Y_{l\hspace{.1cm}hk} + \upsilon,
\end{equation}
be the initial vorticity, for some $k,l$ given, where $\upsilon\in\Dd_0(k')$, for some $k'$.
$\upsilon$ is a perturbation that we assume, without loss of generality, with zero angular momentum.
Let us call $M:=C\cdot f $ and $R:=\sum_{h\in\Zz,|hk|\leq l} \omega^{l\hspace{.1cm}hk}Y_{l\hspace{.1cm}hk}$.
Then, the perturbation $\upsilon$ satisfies:
\[
\dot{\upsilon}=\lbrace\Delta^{-1}(\upsilon+R),(\upsilon+R)\rbrace + \lbrace\Delta^{-1}M,\upsilon+R\rbrace + \lbrace\Delta^{-1}(\upsilon+R),M\rbrace.
\]
The second term on the right hand side can be taken away via the same calculations of Section~\ref{subsec:Coriolis_force} and so we can study:
\begin{equation}\label{eq:pert_simpl2}
\dot{\widetilde{\upsilon}}=\lbrace\Delta^{-1}(\widetilde{\upsilon}+\widetilde{R}),(\widetilde{\upsilon}+\widetilde{R})\rbrace + \lbrace\Delta^{-1}(\widetilde{\upsilon}+\widetilde{R}),M\rbrace,
\end{equation}
where $(\widetilde{\upsilon}+\widetilde{R})(\phi,\theta)=(\Upsilon+R)(\phi-tM/2,\theta)$.
By the Lemma~\ref{lem:RH_interaction2}, the second term on the right hand side of \eqref{eq:pert_simpl2} is such that if $\widetilde{\upsilon}+\widetilde{R}\in\Dd_0(k'')$, for some $k''$, then $\lbrace\Delta^{-1}(\widetilde{\upsilon}+\widetilde{R}),M\rbrace\in\Dd_0(k'')$.
Applying Lemma~\ref{lem:RH_interaction2} to the first term on the right hand side of \eqref{eq:pert_simpl}, we get that $\lbrace\Delta^{-1}(\upsilon+R),(\upsilon+R)\rbrace\in \Dd_0(gcd(k,k'))$.
In conclusion, the vorticity $\omega$ evolves in $\Dd_0(gcd(k,k'))$.
In particular if $\omega(0)$ is the sum of two quantized Rossby-Haurwitz waves of the form \eqref{eq:RH_waves_Dnk}, they evolve in some space $\Dd_0(k'')$, accordingly to the Lemma~\ref{lem:RH_interaction2}.

Finally, we conclude that analogously to the quantized case discussed in Section~\ref{sec:Pert_quant_RH_waves}, we claim that the generic long-time behaviour of a perturbed Rossby-Haurwitz wave in the setting above is such that it leads the vorticity into a quantized quasi-zonal flow.
Indeed, from the inequalities \eqref{eq:cascade_ineq1}, \eqref{eq:cascade_ineq2},  we have that in the energy accumulates at the low frequencies, while the enstrophy is drifted into the high frequencies due to the non-linearity of the Euler equations. 
In particular, in the spaces $\Dd_0(k)$, for $k>2$, the larger the $k$ is, the more the flow will be zonal, since for the lowest possible $l$, the only admissible $m$, multiple of $k$, is $m=0$.
Finally, the number $k$ determines the blobs trapped into these zonal bands.
Indeed, the lowest non-zero non-zonal components correspond to the spherical harmonics $Y_{lk}$, for $l=k,\dots,N$ which have $k$ latitudinal blobs.

\section{Numerical simulations}\label{sec:num_sim}

\begin{figure}[h!]
\centering
\subfloat[Vorticity field $W(t_{0})$]{
  \includegraphics[width=.5\textwidth]{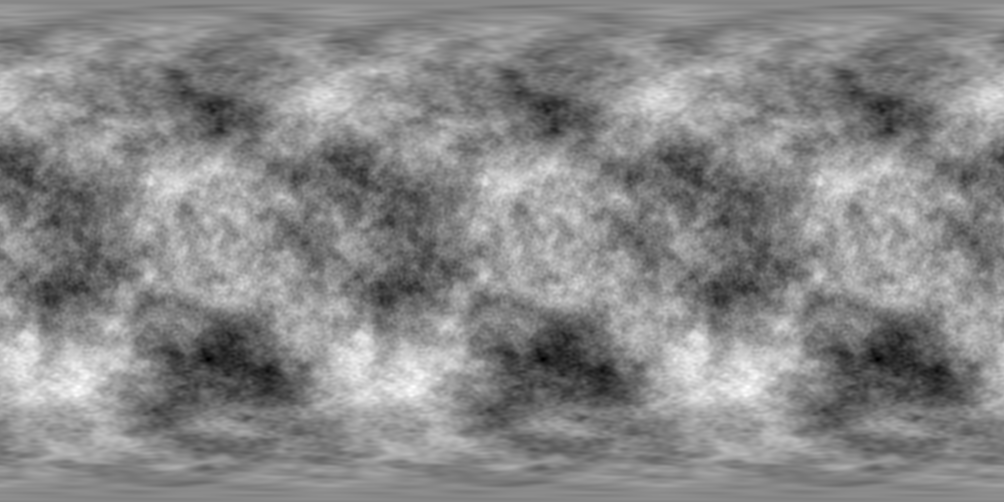}}
\subfloat[Vorticity field $W(t_{end})$]{
  \includegraphics[width=.5\textwidth]{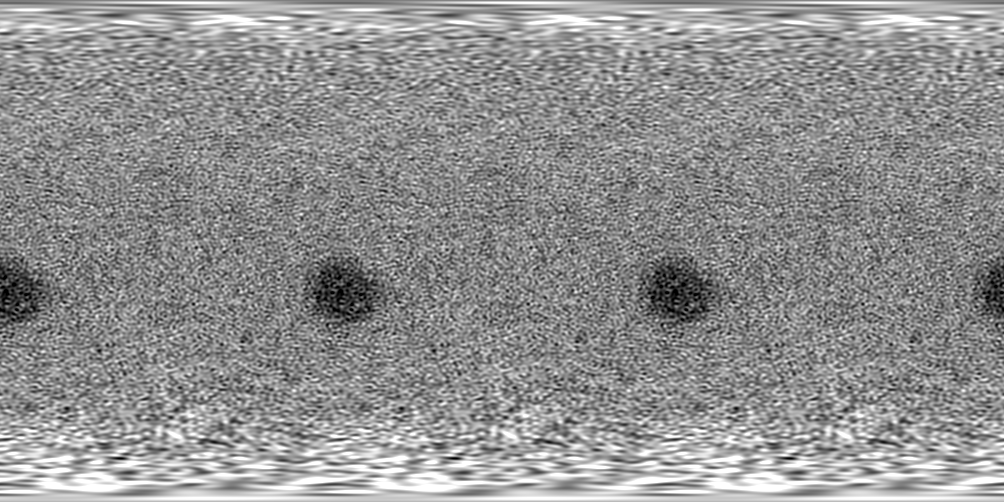}}\\
  \subfloat[Vorticity field $W_s(t_{0})$]{
  \includegraphics[width=.5\textwidth]{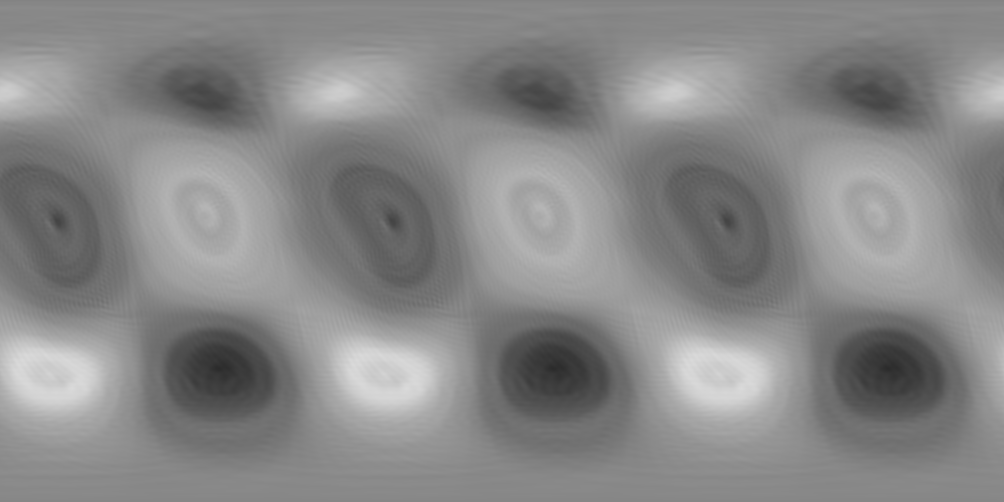}}
\subfloat[Vorticity field $W_s(t_{end})$]{
  \includegraphics[width=.5\textwidth]{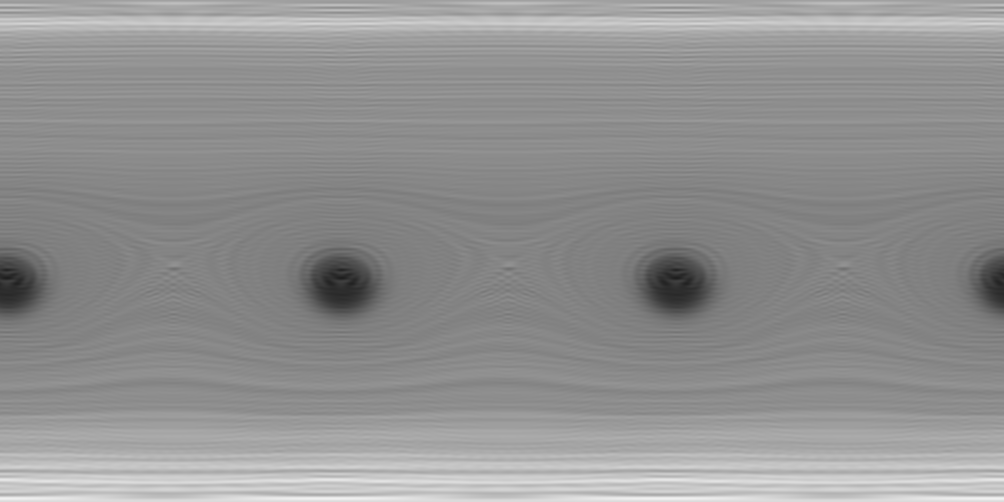}}\\
\caption{The vorticity fields $W,W_s$ at $t=t_0$ and $t=t_{end}$, in $\Dd(257,3)$.}\label{fig:k=3}
\end{figure}

\begin{figure}[h!]
\centering
\subfloat[Vorticity field $W(t_{0})$]{
  \includegraphics[width=.5\textwidth]{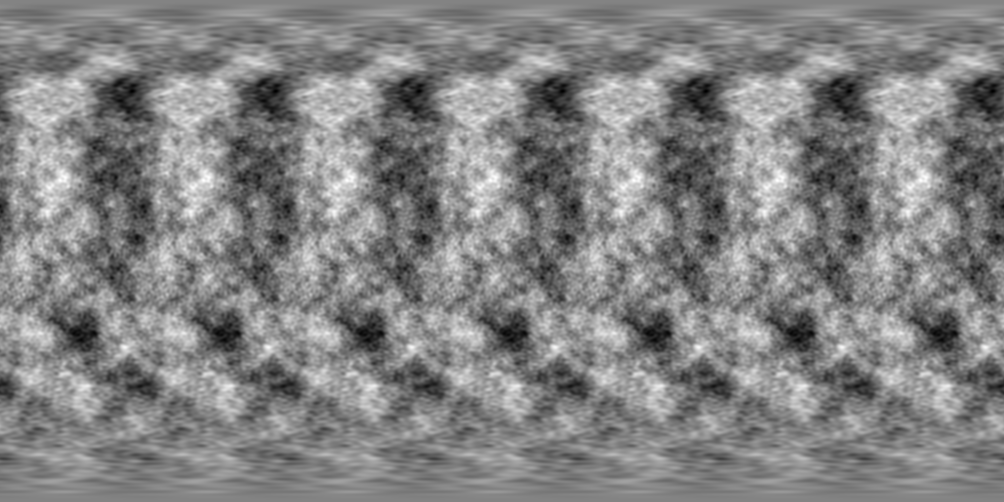}}
\subfloat[Vorticity field $W(t_{end})$]{
  \includegraphics[width=.5\textwidth]{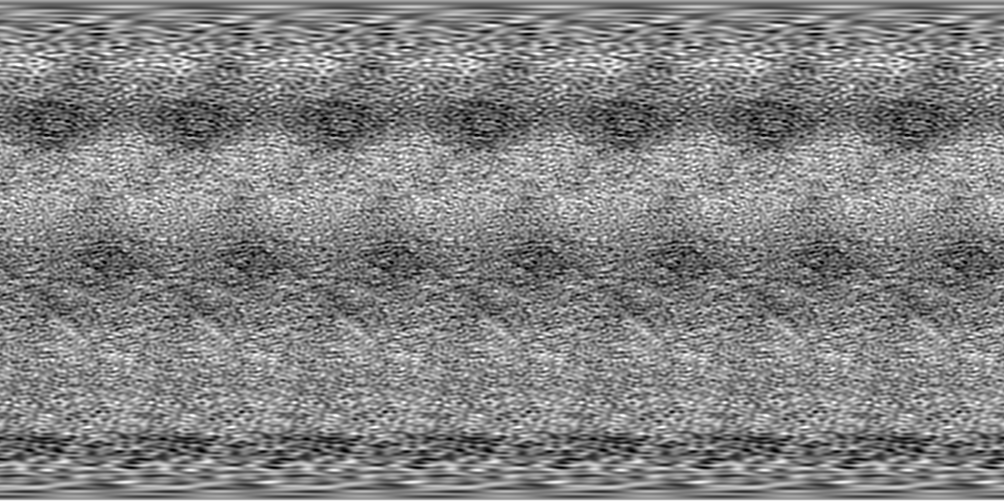}}\\
  \subfloat[Vorticity field $W_s(t_{0})$]{
  \includegraphics[width=.5\textwidth]{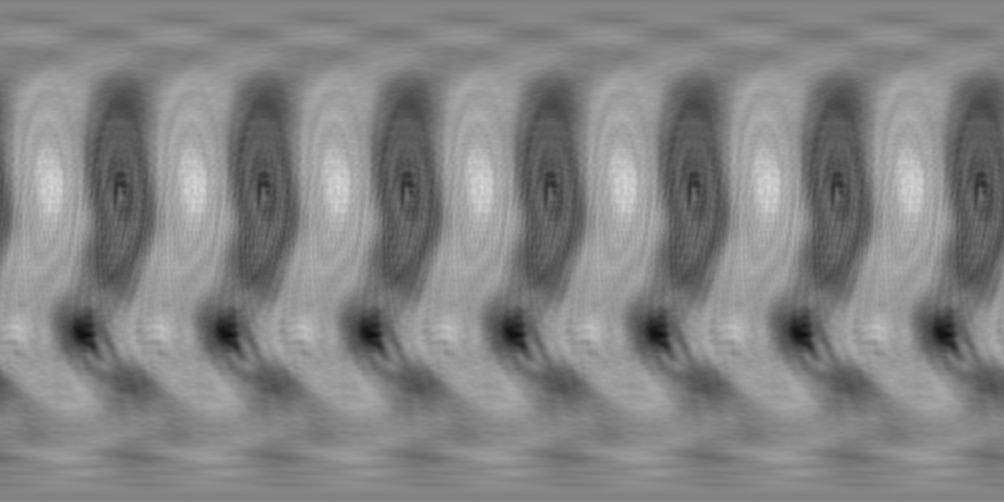}}
\subfloat[Vorticity field $W_s(t_{end})$]{
  \includegraphics[width=.5\textwidth]{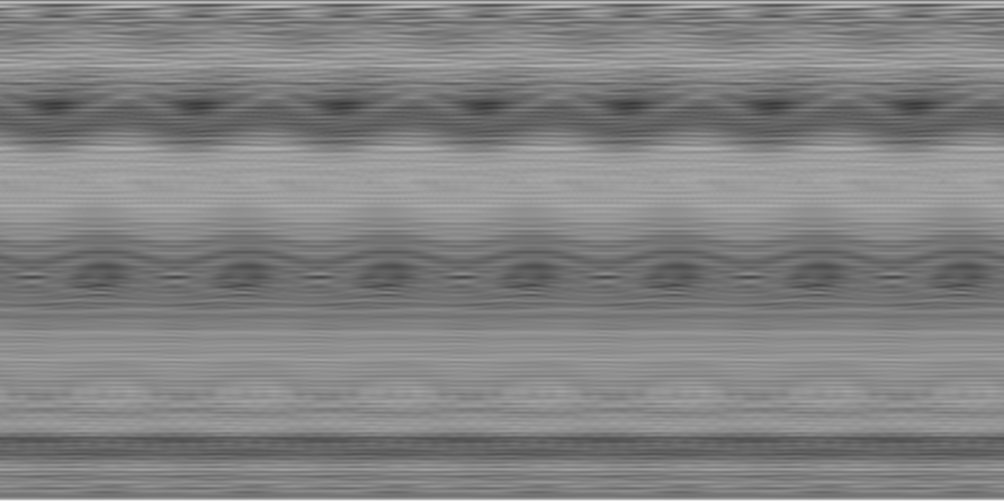}}\\
\caption{The vorticity fields $W,W_s$ at $t=t_0$ and $t=t_{end}$, in $\Dd(257,7)$.}\label{fig:k=7}
\end{figure}

\begin{figure}[h!]
\centering
\subfloat[Vorticity field $W(t_{0})$]{
  \includegraphics[width=.5\textwidth]{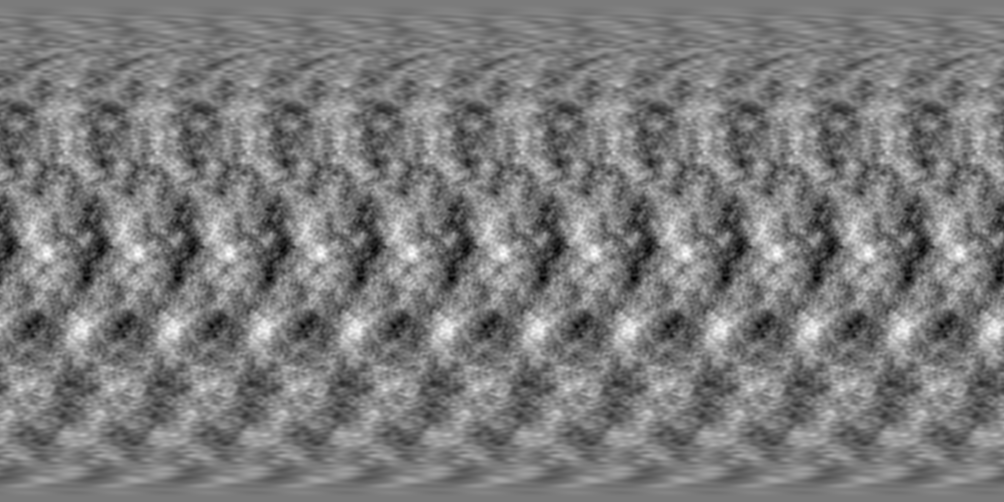}}
\subfloat[Vorticity field $W(t_{end})$]{
  \includegraphics[width=.5\textwidth]{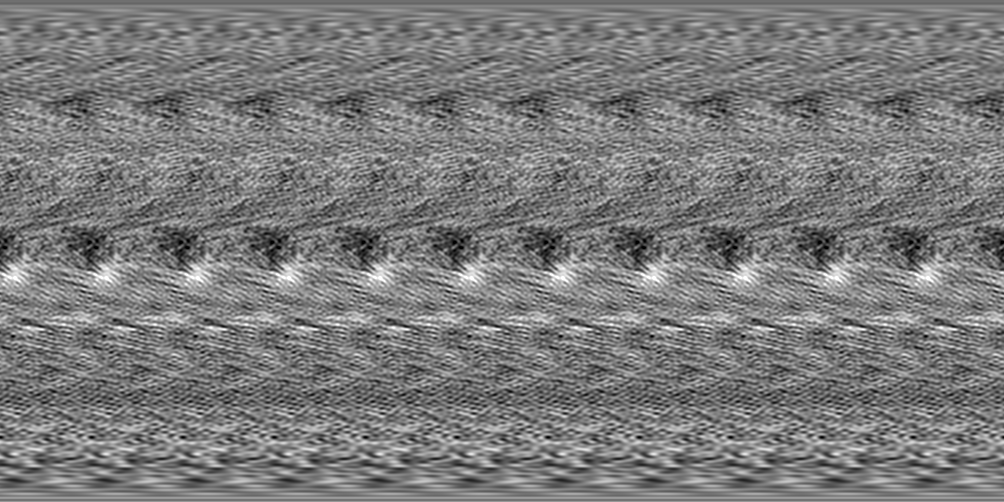}}\\
  \subfloat[Vorticity field $W_s(t_{0})$]{
  \includegraphics[width=.5\textwidth]{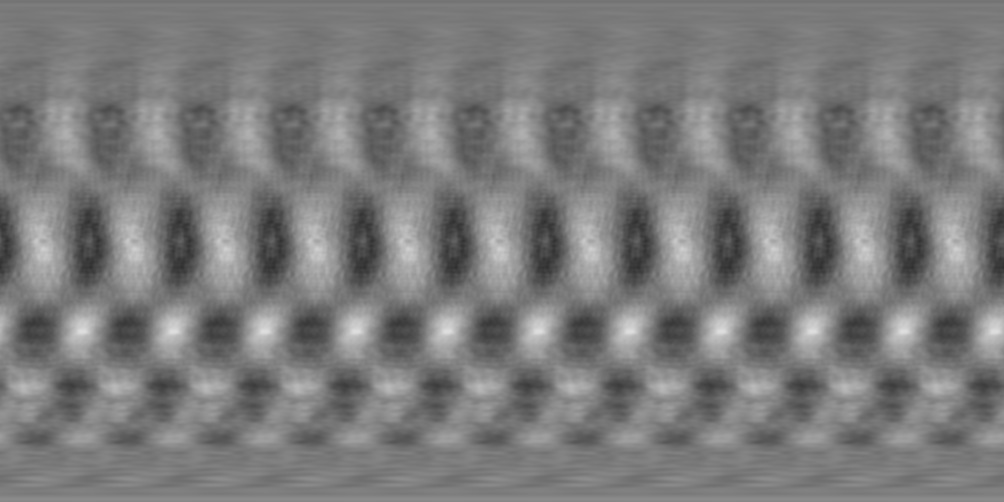}}
\subfloat[Vorticity field $W_s(t_{end})$]{
  \includegraphics[width=.5\textwidth]{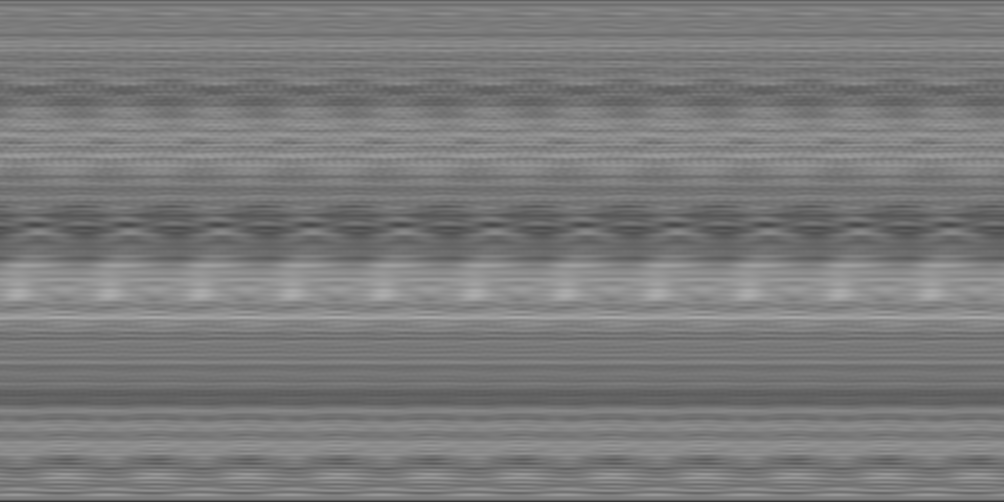}}\\
\caption{The vorticity fields $W,W_s$ at $t=t_0$ and $t=t_{end}$, in $\Dd(257,11)$.}\label{fig:k=11}
\end{figure}

\begin{figure}[h!]
\centering
\subfloat[Vorticity field $W(t_{0})$]{
  \includegraphics[width=.5\textwidth]{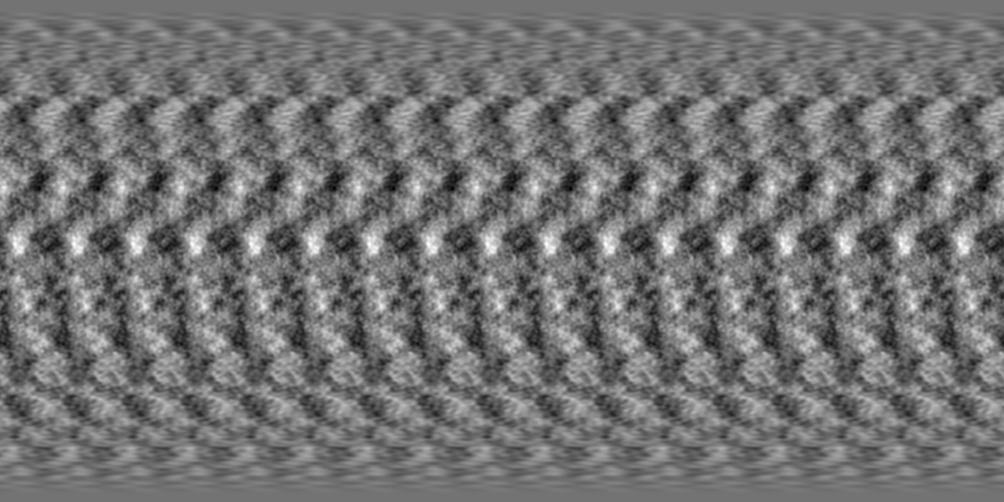}}
\subfloat[Vorticity field $W(t_{end})$]{
  \includegraphics[width=.5\textwidth]{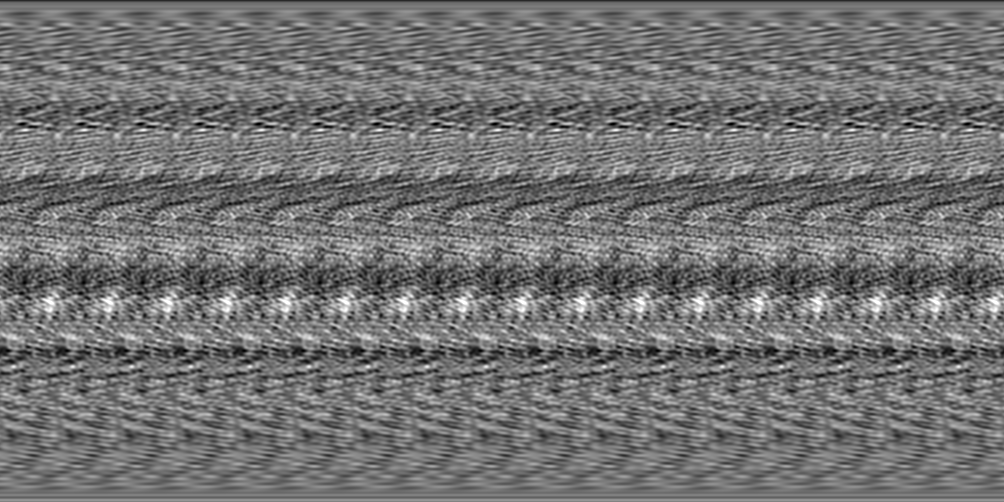}}\\
  \subfloat[Vorticity field $W_s(t_{0})$]{
  \includegraphics[width=.5\textwidth]{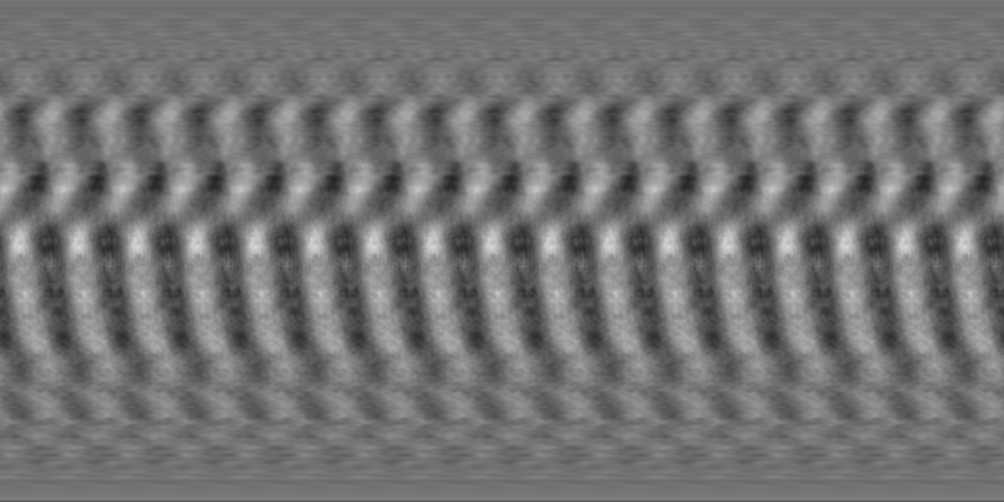}}
\subfloat[Vorticity field $W_s(t_{end})$]{
  \includegraphics[width=.5\textwidth]{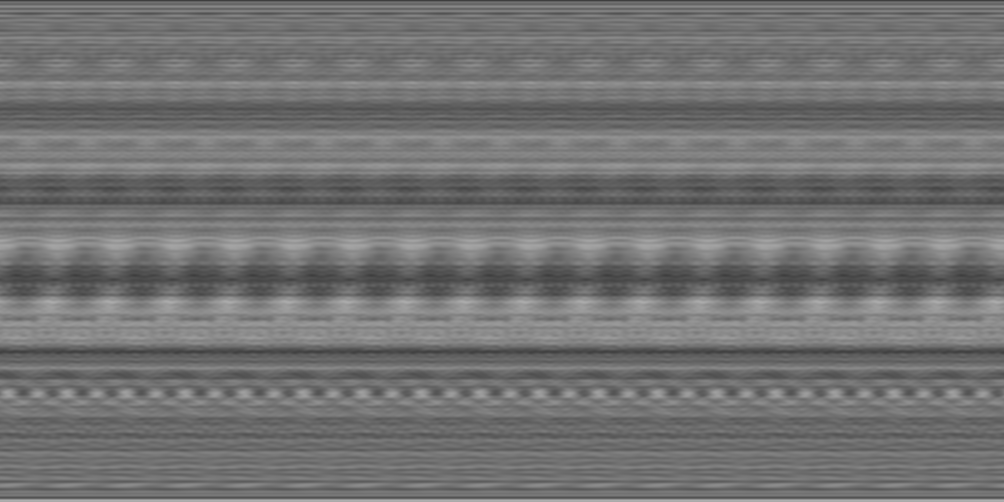}}\\
\caption{The vorticity fields $W,W_s$ at $t=t_0$ and $t=t_{end}$, in $\Dd(257,17)$.}\label{fig:k=17}
\end{figure}

In this section, we present various examples of the results shown above.
More precisely, we integrate equations \eqref{eq:Euler_quant_no_Cor} with the isospectral midpoint method \cite{Viv2019}, for randomly generated\footnote{More specifically, we randomly generate the Fourier components accordingly to the law $W^{lm}l\sim\mathcal{N}(0,1)$.} initial conditions with zero-momentum $\textbf M$ and zero-diagonal components in $\Dd_0(N,k)$, for $N=257$ and different $k$.
We always take time-step $h=0.1$ and normalized initial vorticity, with respect to the spectral norm.
\begin{remark}
The canonical scale separation proposed in \cite{MoVi2021} can be restricted to $\Dd_0(N,k)$.
Indeed, let $W\in\Dd_0(N,k)$, for some $N\geq 1, k\leq N$, and let $W_s$ be the projection of $W$ onto 
\[\mbox{stab}_{\Delta_N^{-1}W}:=\lbrace A\in\SU(N) \mbox{ s.t. } [\Delta_N^{-1}W,A]=0 \rbrace.\]
Then, $W_s\in \Dd_0(N,k)$.
Indeed, any $A\in \mbox{stab}_{\Delta_N^{-1}W}$ can be written as:
\[
A=\sum_{l=1}^N a_l(\Delta_N^{-1}W)^l,
\] and since any power of $\Delta_N^{-1}W$ is in $\Dd_0(N,k)$, also $W_s$ is.
We see this fact in Figures~\ref{fig:k=3},\ref{fig:k=7},\ref{fig:k=11},\ref{fig:k=17}.
\end{remark}
In the plots below, we show in the standard azimuthal-elevation coordinates the vorticity $W$ and its projection $W_s$ at $t=0$ and at $t_{end}$ corresponding to $10^5$ iterations.
As expected from the spherical harmonics symmetries, the vorticity fields in $\Dd_0(N,k)$ have an internal discrete symmetry of the group $\mathbb{Z}/k\mathbb{Z}$.
In the long-times regime, we notice the appearance of jets and $k$ intra-zonal blobs.
We notice that in the projected vorticity field $W_s$, the bands are neater compared to the total vorticity $W$.
In all the simulations, the vorticity at $t=t_0$ is zero on the main diagonal. 
However, at $t=t_{end}$, most of the energy concentrates on the main diagonal components $D:=\mbox{diag}(W)$, see Figure~\ref{tab:Ed_vs_End}.
\begin{figure}[h!]
\begin{tabular}{ |p{4cm}||p{2.5cm}|p{2.5cm}|p{2.5cm}|  }
 \hline
 \multicolumn{4}{|c|}{$H(D(t))/H(W(t))$} \\
 \hline
 Simulation space/Time & $t=t_0$ & $t=(t_{end}+t_0)/2$ & $t=t_{end}$\\
 \hline
 $\Dd(257,3)$  & 0 & 0.8627 & 0.8776\\
 $\Dd(257,7)$  & 0 & 0.9075 & 0.9170\\
 $\Dd(257,11)$ & 0 & 0.6281 & 0.8391\\
 $\Dd(257,17)$ & 0 & 0.6947 & 0.8202\\
 \hline
\end{tabular}
\caption{Ratio of the energy of the diagonal components and total energy  of $W$ at $t=t_0,(t_{end}+t_0)/2,t_{end}$. This values clearly show the evolution of the fluid into a quasi-zonal state.}\label{tab:Ed_vs_End}		
\end{figure}
This fact has been explained in Section~\ref{sec:Pert_quant_RH_waves} and shows that for long times the evolution in the spaces $\Dd(N,k)$ leads the fluid towards a quasi-zonal flow.
The mechanism of concentration of the energy at the lowest possible scales is determined by the inequalities \eqref{eq:cascade_ineq1} and \eqref{eq:cascade_ineq2}.
In Figure~\ref{tab:ineq1} and \ref{tab:ineq2}, we show how these quantities involved in those inequalities evolve during time for the same simulation of Figure~\ref{fig:k=3}.
\begin{figure}[h!]
\begin{tabular}{ |p{2cm}||p{1.5cm}|p{1.5cm}|p{1.5cm}|p{1.5cm}|p{1.5cm}|}
 \hline
Time & $c_H$ & $c_E$ & $\sqrt{\frac{E}{2H}}$ & $c_Hc_E$ & $\frac{E}{2H}$\\
 \hline
$t=t_0$ & 5.1233 & 13.9332 & 6.2284 & 71.3842 & 38.7927 \\
$t=\frac{t_{end}+t_0}{2}$ & 2.7504 & 111.2858 & 6.2284 & 306.0824 & 38.7927\\
$t=t_{end}$ & 2.7116 & 116.3747 & 6.2284 & 315.5591 & 38.7927\\
 \hline
\end{tabular}
\caption{Values of the variables appearing in the inequalities \eqref{eq:cascade_ineq1} at $t=t_0,(t_{end}+t_0)/2,t_{end}$, for the same simulation of Figure~\ref{fig:k=3}.}\label{tab:ineq1}	
\end{figure}

\begin{figure}[h!]
\begin{tabular}{|p{2cm}||p{1.5cm}|p{1.5cm}|p{1.5cm}|p{1.5cm}|p{1.5cm}|}
 \hline
Time & $v_H$ & $v_E$ & $\sqrt{\frac{2H}{E}}$ & $v_Hv_E$ & $\frac{2H}{E}$\\
 \hline
 $t=t_0$ & 0.2169 & 0.1578 & 0.1606 & 0.0342 & 0.0258
\\
  $t=\frac{t_{end}+t_0}{2}$ & 0.4468 & 0.0967 & 0.1606 & 0.0432 & 0.0258\\
   $t=t_{end}$ & 0.4501 & 0.0957 & 0.1606 & 0.0431 & 0.0258\\
 \hline
\end{tabular}
\caption{Values of the variables appearing in the inequalities \eqref{eq:cascade_ineq2} at $t=t_0,(t_{end}+t_0)/2,t_{end}$, for the same simulation of Figure~\ref{fig:k=3}.}\label{tab:ineq2}		
\end{figure}
We notice that the inequalities \eqref{eq:cascade_ineq1} and \eqref{eq:cascade_ineq2} are far from getting sharp, while the time advances.
This indicates one the one hand that the energy concentrates at lower frequencies than those given by the bounds, and on the other hand that the enstrophy concentrates at higher frequencies than the those given by the bounds.

Finally, combining Figure~\ref{tab:Ed_vs_End} with Figure~\ref{tab:ineq1} and \ref{tab:ineq2}, we see that as expected the energy must concentrate at the lowest accessible diagonal modes, due by the combined effect of the inverse energy cascade and the algebraic structure of the spaces $\Dd(N,k)$, as discussed in Section~\ref{sec:Pert_quant_RH_waves}.

\newpage
Finally, we show the result of Theorem~\ref{thm:comp_adj} and Remark~\ref{rem:split_dyn} for the final and initial vorticity of the same simulation of Figure~\ref{fig:k=3}.
For this simulation we have $\Dd_0(257,3) \cong \mathfrak{s}(\UU\left(85\right)\oplus \UU\left(86\right)^2)$.

%For acknowledgements section, please don't number the section, please begin it with \section*{Acknowledgements}
\section*{Acknowledgments} The author thanks prof. Klas Modin for the enlightening discussions and support to complete this work.

% You may incorporate your references as follows in your main tex file.
% Using BibTex is not recommended but can be handled.
%\begin{thebibliography}{99}

\bibliographystyle{plainnat}
%\IfFileExists{/Users/moklas/Documents/Papers/References.bib}{
%\bibliography{/Users/moklas/Documents/Papers/References}
%}{
\bibliography{biblio}

\end{document}